\documentclass[a4paper,11pt]{article}
\usepackage{jheppub} 
\usepackage[utf8]{inputenc}
\usepackage{physics}
\usepackage{slashed}
\usepackage{caption}
\usepackage{xcolor}
\usepackage{comment}
\usepackage{multirow}
\usepackage{graphics}
\usepackage{float}
\usepackage{cases}
\usepackage{cancel}
\usepackage{soul}
\usepackage{array}
\usepackage{mathtools}  
\usepackage{amsfonts}
\usepackage{hyperref}
\usepackage{amsmath}
\usepackage{amssymb}
\usepackage{tcolorbox}
\usepackage{tikz}
\usetikzlibrary{arrows.meta, positioning}

\title{\boldmath Super-Penrose $\And$ Witten Transforms for SCFT$_3$}

\author{Deep Mazumdar}

\affiliation{Indian Institute of Science Education and Research,\\ Dr Homi Bhabha Road, Pashan, Pune, India}

\emailAdd{deepkamal.mazumdar@students.iiserpune.ac.in}

\abstract{The study of three dimensional CFT correlators in twistor space has recently garnered a significant interest. Conformal symmetry acts linearly in the twistor space, which streamlines the analysis. Moreover, twistors provide a connection to the position and momentum space through the Penrose and Witten transforms, respectively. In this work, we develop the supersymmetric versions of Penrose and Witten transforms for three dimensional superconformal field theories for $\mathcal{N}=1\;\textrm{to}\;4$. We derive two and three point functions in the position and momentum superspace for the $\mathcal{N}=1$ scenario using these transforms. Extending this setup to higher supersymmetries turns out to be a simple extension of the $\mathcal{N}=1$ case, which aligns with the inherent simplicity of supertwistors. 
}

\begin{document}
\maketitle

\section{Introduction}
Supersymmetry plays a crucial role as a theoretical framework. Theories like $\mathcal{N}= 4$ super Yang-Mills theory in four dimensions serve as a mathematical laboratory and have paved the way to understand several physical aspects. The study of superamplitudes \cite{Elvang:2013cua} has seen remarkable progress through the use of modern tools such as supertwistors \cite{Ferber:1977qx,Witten:2003nn}, Grassmannian geometry \cite{Arkani-Hamed:2012zlh} and amplitudhedron \cite{Arkani-Hamed:2013jha}.

Another avenue with significant physical and mathematical richness lies in the study of conformal field theories (CFTs). Traditionally, CFT correlators were computed in the position space \cite{Belavin:1984vu,Giombi:2011rz,Rattazzi:2008pe}, and later using the embedding space formalism \cite{Costa:2011mg,Costa:2011dw}. However, studying them in momentum space \cite{McFadden:2011kk,Ghosh:2014kba,Coriano:2013jba,Bzowski:2013sza,Bzowski:2015pba,Bzowski:2017poo,Farrow:2018yni,Bzowski:2018fql,Skvortsov:2018uru,Bautista:2019qxj,Lipstein:2019mpu,Isono:2019ihz,Gillioz:2019lgs,Baumann:2019oyu,Jain:2020rmw,Jain:2020puw,Jain:2021wyn,Jain:2021qcl,Caron-Huot:2021kjy,Jain:2021gwa,Jain:2021whr} has led to several interesting results that are not obvious from the position space perspective. In recent years, substantial computational advances have been achieved in the analysis of three dimensional CFT correlators, driven by modern amplitude-inspired techniques involving spinor-helicity \cite{Maldacena:2011nz,Baumann:2020dch,Baumann:2021fxj,Jain:2021vrv,Jain:2024bza} and twistors \cite{Baumann:2024ttn,Bala:2025gmz,Bala:2025jbh,Bala:2025qxr}.

Superconformal field theories (SCFTs) present a compelling synthesis of supersymmetry and conformal symmetry. In three dimensions, models such as the $\mathcal{N}=6$ ABJM theory \cite{Aharony:2008ug} are of paramount physical importance. SCFT correlators are typically studied in position superspace \cite{Osborn:1998qu,Park:1999pd,Park:1999cw,Nizami:2013tpa,Buchbinder:2015qsa,Buchbinder:2015wia,Kuzenko:2016cmf,Buchbinder:2021gwu,Buchbinder:2021izb,Buchbinder:2021kjk,Buchbinder:2021qlb,Jain:2022izp,Buchbinder:2023fqv,Buchbinder:2023ndg}, where supercorrelators are built from superconformally invariant building blocks. While this framework has many advantages, it becomes technically challenging: especially in theories with extended supersymmetry. The main difficulty arises from the proliferation of independent tensor structures, which complicates the analysis. In that vein, some efforts have been made to alleviate this issue by exploring amplitude-inspired techniques like supertwistors as embedding space spinors \cite{Kuzenko:2010rp,Kuzenko:2012tb,deAzcarraga:2014hda}, Grassmann twistors \cite{Jain:2023idr}, and ultimately a manifest twistor superspace formalism \cite{Bala:2025jbh}, offering a more streamlined framework for handling supersymmetric correlators.

In this work, we aim to establish the relationship between position superspace, momentum superspace and twistor superspace for supercorrelators in three dimensional superconformal field theories with varying degrees of supersymmetry, as schematically illustrated in figure \ref{fig:1}. While a direct Fourier transform between position and momentum superspace is challenging, we achieve this connection by developing the supersymmetric versions of the Penrose and Witten transform. The super-Penrose transform maps results from twistor superspace to their position superspace analogue. Using this framework, we derive two and three point functions by performing projective integrals in a fully covariant manner. Along the way, we discuss several important subtleties and technical nuances involved in these computations. Moreover, this method evades all the difficulties associated with the increasing number of invariant blocks. We also obtain the corresponding momentum superspace expressions in spinor-helicity variables by applying the super-Witten transform to the twistor superspace counterparts.
\begin{center}\label{fig:1}
\begin{tikzpicture}[
    node distance=2.85cm and 2.85cm,
    box/.style={rectangle, draw, rounded corners, align=center, minimum width=2.85cm, minimum height=1cm},
    arrow/.style={-{Latex[length=3mm, width=2mm]}, thick},
]
\node[box] (T) {Supertwistors};
\node[box, below left=of T] (P) {Position\\superspace};
\node[box, below right=of T] (M) {Momentum\\superspace};
\draw[->, thick] (T.south west) -- (P.north)
node[midway, above, sloped] {Super-Penrose};
\draw[->, thick] (T.south east) -- (M.north)
node[midway, above, sloped] {Super-Witten};
\draw[<->, thick] (P) -- (M)
node[midway, above] {Super-Fourier};
\end{tikzpicture}
\newline
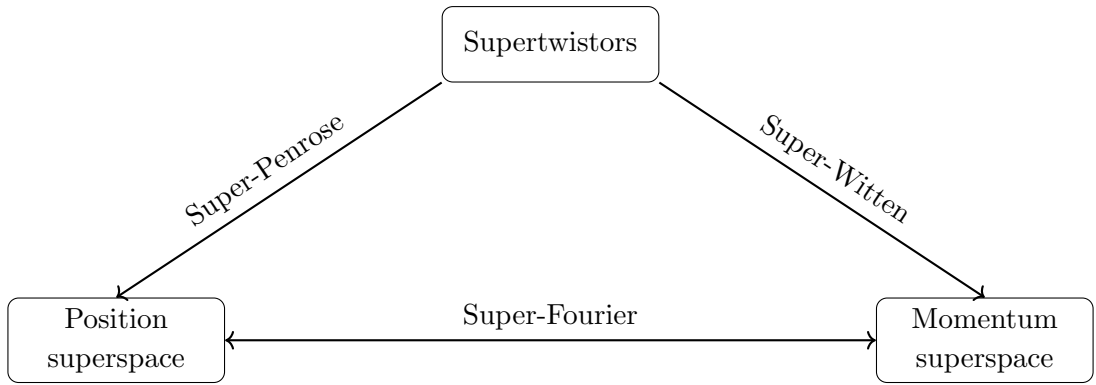
\textbf{Figure 1:}  The network of twistor, position and momentum superspaces.
\end{center}  

\subsection*{Outline}
A detailed outline of the paper is as follows: In section \ref{sec:2}, we present a brief review of the Penrose and Witten transforms in three dimensions for the non-supersymmetric setting. Thereafter, we develop the super-Penrose transform in section \ref{sec:3} and obtain the position superspace two and three point functions. Following up, we develop the super-Witten transform in section \ref{sec:4} to obtain the momentum superspace results for two and three point functions in spinor-helicity variables. We then develop the aforementioned transforms for cases involving extended supersymmetric in section \ref{sec:5}. Finally, in section \ref{Discussion}, we summarize this work and discuss a number of interesting future directions.

We supplement this paper with a few essential appendices. We layout the notation and conventions followed throughout this work in appendix \ref{app:1}. We then provide some useful position superspace identities in appendix \ref{app:2} and necessary spinor-helicity identities in appendix \ref{app:3} respectively. We present a detailed example for the super-Penrose transform in appendix \ref{app:4}, followed by an example for the super-Witten transform in appendix \ref{app:5}.

\section{A prelude to twistor space}\label{sec:2}
In this section, we present a short review of the twistor space. Originally formulated in \cite{Penrose:1967wn} for four spacetime dimensions, a similar setup was recently developed for three dimensions in \cite{Baumann:2024ttn,Bala:2025gmz} to study CFT correlators in twistor space. In \cite{Baumann:2024ttn,Bala:2025gmz}, the twistors $Z=(\lambda,\bar{\mu})$ were derived starting from the spinor-helicity variables ($\lambda,\bar{\lambda}$) by performing a half-Fourier transform, i.e. a Fourier transform with respect to either $\lambda$ or $\bar{\lambda}$
\begin{align}
(\lambda,\bar{\lambda})\longrightarrow (\lambda,\bar{\mu}).
\end{align}
The twistors ($Z$) not only simplify the computation of correlation functions, but also transform naturally under the spin representation of three dimensional conformal group Sp$(4;\mathbb{R})$, thus streamlining the analysis.

Furthermore, one can map correlators from the twistor space to their position/momentum space counterparts using integral transforms, as alluded in \cite{Baumann:2024ttn,Bala:2025gmz}.
In this section, we briefly discuss the three dimensional versions of Penrose \cite{Penrose:1967wn,Penrose:1969ae} and Witten transforms\footnote{Technically, this is the inverse Witten transform, but we shall refer to it as the Witten transform for brevity.} \cite{Witten:2003nn} of functions in twistor space. While the former leads to position space results, the latter leads to its momentum space avatar. We discuss these transforms, starting with the Penrose transform first.

\subsection{Penrose transform}
The Penrose transform provides a bridge from the twistor space to the position space
\begin{align}
(\lambda,\bar{\mu})\longrightarrow{\textrm{}}x^\mu.
\end{align}
From the field theory perspective, it translates holomorphic structures in the twistor space to massless field solutions in position space \cite{Penrose:1969ae}.

The Penrose transform is a projective integral of $\lambda$ over $\mathbb{RP}^1$, subject to the incidence relation: $\bar{\mu}^a=x^{a}_b\lambda^b$. For a function $\hat{f}_s(\lambda,\bar{\mu})$ of spin `s', it is defined as 
\begin{align}\label{Penrose}
f_s^{a_1\cdots a_{2s}}(x)=\int_{\mathbb{RP}^1}D\lambda \prod_{i=1}^{2s}\lambda^{a_i}\;\hat{f}_s(\lambda,\bar{\mu})\bigg\rvert_{\bar{\mu}^a=x^{a}_b\lambda^b},
\end{align}
where the projective measure is defined as $D\lambda=\lambda\wedge d\lambda$.
Under projective rescaling $\lambda\rightarrow r\lambda$, the measure scales by $r^2$. The rescaling invariance of $f_s^{a_1\cdots a_{2s}}(x)$ demands that the function $\hat{f}_s(\lambda,\bar{\mu})$ scale as $r^{2s+2}$.

Let us now illustrate the Penrose transform explicitly for CFT correlation functions. Consider the two point function of conserved currents of spin `s' in the twistor space \cite{Baumann:2024ttn,Bala:2025gmz}
\begin{align}\label{2ptCFT}
\langle0|\hat{J}_s(\bar{\mu}_1,\lambda_{1})\hat{J}_s(\bar{\mu}_2,\lambda_{2})|0\rangle=\frac{i^{2s+2}}{(\lambda_1\cdot\bar{\mu}_2-\lambda_2\cdot\bar{\mu}_1)^{2s+2}},
\end{align}
where the spinor indices are contracted from bottom to top i.e. $\lambda_i\cdot\bar{\mu}_j\equiv\lambda_{i,a}\bar{\mu}_j^a$.

The corresponding Penrose transform for \eqref{2ptCFT} is
\begin{align}\label{2ptCFT1}
\langle0|J_s^{a_1\dots a_{2s}}(x_{1})J_s^{b_1\dots b_{2s}}(x_2))|0\rangle=&\int_{\mathbb{RP}^1}D\lambda_1D\lambda_2\prod_{i=1}^{2s}\lambda_1^{a_i}\lambda_2^{b_i}\;\frac{i^{2s+2}}{(\lambda_{1a}(x_2)_b^{a}\lambda_{2}^b-\lambda_{2a}(x_1)_b^{a}\lambda_{1}^b)^{2s+2}}\notag\\
=&\int_{\mathbb{RP}^1}D\lambda_1D\lambda_2 \prod_{i=1}^{2s}\lambda_1^{a_i}\lambda_2^{b_i}\;\frac{i^{2s+2}}{(\lambda_{1a}(x_{12})_{b}^{a}\lambda_{2}^b)^{2s+2}}.
\end{align}
The integrand in \eqref{2ptCFT1} can be expressed as a Schwinger integral
\begin{align}\label{2ptCFT2}
\langle0|J_s^{a_1\dots a_{2s}}(x_{1})J_s^{b_1\dots b_{2s}}(x_2))|0\rangle=&\int_{\mathbb{RP}^1}D\lambda_1D\lambda_2 \prod_{i=1}^{2s}\lambda_1^{a_i}\lambda_2^{b_i}\int dc_{12}\;c_{12}^{2s+1}\textrm{Sgn}(c_{12})\textrm{exp}(ic_{12}\lambda_{1a}(x_{12})_{b}^{a}\lambda_{2}^b).
\end{align}
Contracting with polarization spinors $\xi_i$, \eqref{2ptCFT2} is solved using partial integrals
\begin{align}\label{2ptCFTmid}
&\langle0|J_s(x_{1})J_s(x_2))|0\rangle\notag\\
&=\int_{\mathbb{RP}^1}D\lambda_1D\lambda_2\; (\xi_1\cdot\lambda_1)^{2s}(\xi_2\cdot\lambda_2)^{2s}\int dc_{12}\;c_{12}^{2s+1}\textrm{Sgn}(c_{12})\textrm{exp}(-ic_{12}\lambda_{1a}\lambda_{2b}(x_{12})^{ab})\notag\\
&=\int_{\mathbb{RP}^1}D\lambda_1D\lambda_2 (\xi_1\cdot\lambda_1)^{2s}\Big(\frac{\xi_{2m}(x_{12})^{mn}}{x_{12}^2}\frac{\partial}{\partial\lambda_1^n}\Big)^{2s}\int dc_{12}\;c_{12}\textrm{Sgn}(c_{12})\textrm{exp}(-ic_{12}\lambda_{1a}\lambda_{2b}(x_{12})^{ab})\notag\\
&=\Big(\frac{\xi_{1n}\xi_{2m}(x_{12})^{mn}}{x_{12}^2}\Big)^{2s}\int_{\mathbb{RP}^1}D\lambda_1D\lambda_2 \int dc_{12}\;\textrm{Sgn}(c_{12})\textrm{exp}(-ic_{12}\lambda_{1a}\lambda_{2b}(x_{12})^{ab}),
\end{align}
where the total derivative term is zero. This is due to the fact that the twistor space correlators are technically distributional in nature, which are well defined only when integrated against Schwartz functions, and have fall-off at boundary value of these integrals.

Combining the projective measure $D\lambda_2$ with the Schwinger integral $dc_{12}$ gives the non-projective measure $d^2\lambda_2$. Then the resultant integral leads to\footnote{We defer the details of this calculation in appendix \ref{app:2}.}
\begin{align}\label{Mod}
&\int_{\mathbb{RP}^1}D\lambda_1D\lambda_2\int dc_{12}\;\textrm{Sgn}(c_{12})\textrm{exp}(-ic_{12}\lambda_{1a}\lambda_{2b}(x_{12})^{ab})=\frac{1}{2x_{12}^2}.
\end{align}
Plugging \eqref{Mod} into \eqref{2ptCFTmid}, we obtain the position space result for two point function
\begin{align}\label{2ptCFTpos}
\langle0|J_s(x_{1})J_s(x_2))|0\rangle=\frac{(\xi_{1a}(x_{12})^{a}_b\xi_{2}^b)^{2s}}{2|x_{12}|^{4s+2}}=\frac{(P_{3})^{2s}}{2|x_{12}|^2},
\end{align}
where the final result is expressed in terms of the inversion invariant conformal building block $P_3$ mentioned in \cite{Giombi:2011rz}.
Let us now briefly discuss the Witten transform.

\subsection{Witten transform}
The Witten transform acts as a map from twistor space to momentum space spinor-helicity variables, via a half-Fourier transform over the variable $\bar{\mu}$
\begin{align}
(\lambda,\bar{\mu})\longrightarrow(\lambda,\bar{\lambda}).
\end{align}
In the context of scattering amplitudes, it establishes a correspondence between geometric objects in twistor space and the kinematics of massless particles \cite{Witten:2003nn}.

Consider a function $\hat{f}_s(\lambda,\bar{\mu})$ of spin `s'. The Witten transform for the three dimensional case is defined as 
\begin{align}\label{Witten}
\tilde{f}_s(\lambda,\bar{\lambda})=\int{d^2\bar{\mu}}\;\textrm{exp}(-i\bar{\lambda}_a\bar{\mu}^a)\hat{f}_s(\lambda,\bar{\mu}).  
\end{align}
Let us illustrate this using a simple example. Consider the two point function of conserved currents of spin `s' in the twistor space \eqref{2ptCFT}.
The Witten transform is as follows
\begin{align}
\langle 0| \tilde{J}_s(\lambda_1,\bar{\lambda}_1)\tilde{J}_s(\lambda_2,\bar{\lambda}_2)| 0\rangle&=\int\prod_{i=1}^2{d^2\bar{\mu}_i}\textrm{exp}(-i\bar{\lambda}_{ia}\bar{\mu}_i^a)\int dc_{12}\;c_{12}^{2s+1}\textrm{Sgn}(c_{12})\textrm{exp}(ic_{12}(\lambda_1\cdot\bar{\mu}_2-\lambda_2\cdot\bar{\mu}_1))\notag\\
&=\int{d^2\bar{\mu}_1}{d^2\bar{\mu}_2}\int dc_{12}c_{12}^{2s+1}\textrm{Sgn}(c_{12})\textrm{exp}(i(\bar{\lambda}_{1a}+c_{12}\lambda_{2a})\bar{\mu}_1^a)\textrm{exp}(i(\bar{\lambda}_{2a}-c_{12}\lambda_{1a})\bar{\mu}_2^a).
\end{align}
Performing the integrations for $\bar{\mu}_i$ results in
\begin{align}\label{2ptWitten2}
&\langle 0| \tilde{J}_s(\lambda_1,\bar{\lambda}_1)\tilde{J}_s(\lambda_2,\bar{\lambda}_2)| 0\rangle=\int dc_{12}c_{12}^{2s+1}\textrm{Sgn}(c_{12})\delta^{(2)}(\bar{\lambda}_{1a}+c_{12}\lambda_{2a})\delta^{(2)}(\bar{\lambda}_{2a}-c_{12}\lambda_{1a}).
\end{align}
We now perform a complete basis change to $\lambda_{1}\And\lambda_{2}$ using the Schouten identities \eqref{Schouten}. Using a simple identity of the delta function \eqref{Decomp}, the expression \eqref{2ptWitten2} can be recast in the following form
\begin{align}\label{2ptWitten3}
\langle 0| \tilde{J}_s(\lambda_1,\bar{\lambda}_1)\tilde{J}_s(\lambda_2,\bar{\lambda}_2)| 0\rangle&=\int dc_{12}\;c_{12}^{2s+1}\textrm{Sgn}(c_{12})\frac{1}{|\langle 12\rangle|^{2}}\delta\left(c_{12}-\frac{\langle\bar{1}1\rangle}{\langle12\rangle}\right)\delta\left(\frac{\langle\bar{2}1\rangle}{\langle12\rangle}\right)\delta\left(\frac{\langle \bar{1}2\rangle}{\langle 12\rangle}\right) \delta\left(c_{12}-\frac{\langle \bar{2}2\rangle}{\langle 12\rangle}\right)\notag\\
&=\frac{1}{|\langle 12\rangle|^{2}}\left(\frac{\langle\bar{1}1\rangle}{\langle12\rangle}\right)^{2s+1}\textrm{Sign}\left(\frac{\langle\bar{1}1\rangle}{\langle12\rangle}\right)\delta\left(\frac{\langle\bar{2}1\rangle}{\langle12\rangle}\right)\delta\left(\frac{\langle \bar{1}2\rangle}{\langle 12\rangle}\right) \delta\left(\frac{\langle\bar{1}1\rangle}{\langle12\rangle}-\frac{\langle \bar{2}2\rangle}{\langle 12\rangle}\right).
\end{align}
Some simple manipulations from here result in the product of a kinematic factor and a momentum-conserving delta function. We urge the reader to refer appendix \ref{app:3} for the details of the same. Thus, the momentum space result for the two point function is\footnote{We suppress the factor of $\textrm{Sign}(\langle\bar11\rangle)$, which is positive for the regime of spacelike momenta \cite{Bala:2025gmz}.}
\begin{align}\label{2ptCFTmom}
\langle 0| \tilde{J}_s(\lambda_1,\bar{\lambda}_1)\tilde{J}_s(\lambda_2,\bar{\lambda}_2)| 0\rangle=\frac{\langle\bar{1}\bar{2}\rangle^{2s}}{(2p_1)^{2s-1}}\delta^{(3)}(\vec{p}_1+\vec{p}_2),
\end{align}
which is proportional to the spinor-helicity result \cite{Maldacena:2011nz}.
Thus far we have observed that the position and momentum space correlators can be obtained from their twistor space counterparts via the Penrose \eqref{Penrose} and Witten \eqref{Witten} transforms, respectively. In the subsequent sections, we will explore this idea for the supersymmetric scenario and establish the supersymmetric analogue of these transforms. Let us begin with the super-Penrose transform first.

\section{Super-Penrose transform for $\mathcal{N}=1$}\label{sec:3}
In the previous section, we saw how the Penrose transform \eqref{Penrose} acts as a bridge from the twistor space to position space. In that vein, the super-Penrose transform is the supersymmetric analogue of the Penrose transform. While this looks as a straightforward generalization at first, the supersymmetric avatar involves several crucial nuances to which its non-supersymmetric version is oblivious. For starters, one needs to deal with the fermionic coordinates appropriately. Thus, the super-Penrose transform takes a very interesting form, as we shall see in the following discussion. 

The super-Penrose transform was first envisioned for four dimensional SCFTs in \cite{Ferber:1977qx}. There, the transform amounts to an integral over bosonic spinors, subject to some appropriate incidence relations for bosonic and fermionic coordinates. We follow a similar approach to develop its three dimensional version with $\mathcal{N}=1$ supersymmetry.

\subsection{Super-Penrose transform: Supertwistors $\rightarrow$ position superspace}
For the three dimensional scenario, the super-Penrose transform maps from twistor superspace $\mathcal{Z}=(\lambda,\bar{\mu},\psi)$ as defined in \cite{Bala:2025jbh} to position superspace
\begin{align}
(\lambda,\bar{\mu},\psi)\longrightarrow(x^\mu,\theta).
\end{align}
Let us first begin by deriving the incidence relations. Consider the supersymmetry invariant interval for the $\mathcal{N}=1$ case \cite{Bala:2025jbh}
\begin{align}\label{Z1Z2}
\mathcal{Z}_1\cdot\mathcal{Z}_2=\lambda_1\cdot\bar{\mu}_2-\lambda_2\cdot\bar{\mu}_1-\psi_1\psi_2.
\end{align}
The supersymmetric incidence relation acts as a bridge from the twistor superspace to position superspace coordinates in the following manner
\begin{align}
\mathfrak{X}:=\qquad\Bigg(\bar{\mu}^a=x^{a}_b\lambda^b+\alpha\;\theta^2\lambda^a,\qquad
\psi=\beta\;\theta^a\lambda_a\Bigg).    
\end{align}
The invariant interval \eqref{Z1Z2} subject to $\mathfrak{X}$ is
\begin{align}\label{Z1Z2X}
\mathcal{Z}_1\cdot\mathcal{Z}_2\bigg\rvert_{\mathfrak{X}}=\lambda_{1a}\lambda_{2b}\Big((x_{12})^{ab}+\alpha(\theta_1^2+\theta_2^2)\epsilon^{ab}-\beta^2\theta_1^a\theta_2^b\Big).
\end{align}
We now demand supersymmetry invariance for \eqref{Z1Z2X} in the position superspace
\begin{align}
\sum_{i=1}^2Q_{i,a}\Big(\mathcal{Z}_1\cdot\mathcal{Z}_2\Big)\bigg\rvert_{\mathfrak{X}}=0,
\end{align}
where the supercharge generator is given by \cite{Jain:2023idr}
\begin{align}
Q_{i,a}=\frac{\partial}{\partial\theta^a}+\frac{i}{2}\theta_{i,b}\;\slashed{\partial}_a^b.
\end{align}
This fixes the coefficients $\alpha$ and $\beta$
\begin{align}\alpha=-\frac{i}{4},\qquad\beta=e^{-\frac{i\pi}{4}}.
\end{align}
Thus, the incidence relations for bosonic and fermionic coordinates $\bar{\mu}\And\psi$ are
\begin{align}\label{IncidenceSUSY}
\mathfrak{X}:=\qquad\Bigg(\bar{\mu}^a=x^{a}_b\lambda^b-\frac{i}{4}\theta^2\lambda^a,\qquad
\psi=e^{-\frac{i\pi}{4}}\theta^a\lambda_a\Bigg).    
\end{align}
Having derived the supersymmetric incidence relation \eqref{IncidenceSUSY}, we define the super-Penrose transform as the projective integral of $\lambda$ over $\mathbb{RP}^1$, subject to \eqref{IncidenceSUSY}. Consider a function $\hat{\mathbf{f}}_s(\lambda,\bar{\mu},\psi)$ of spin `s' in the twistor superspace. The super-Penrose transform is defined as\footnote{A crucial property of the Penrose transform in non supersymmetric context is that the conservation of currents is manifest \cite{Baumann:2024ttn,Bala:2025qxr}. It can be easily checked that the incidence relation \eqref{IncidenceSUSY} here ensures the conservation of supercurrents as well.}
\begin{align}\label{SUSYPenrose}
\mathbf{f}_s^{a_1\cdots a_{2s}}(x,\theta)=\int_{\mathbb{RP}^1}D\lambda \prod_{i=1}^{2s}\lambda^{a_i}\;\hat{\mathbf{f}}_s(\lambda,\bar{\mu},\theta)\bigg\rvert_{\mathfrak{X}}.
\end{align}
With the super-Penrose transform \eqref{SUSYPenrose} in our arsenal, let us derive the correlation functions in position superspace. 
\subsection{Examples}
We start with the simple case of two point functions of conserved supercurrents.
\subsection*{Two point functions}
The two point functions in the twistor superspace are as follows\footnote{Henceforth, we shall suppress the overall factors of $i^{2s+2}$ for two points and $i^{s_1+s_2+s_3}$ for three points.} \cite{Bala:2025jbh}
\begin{align}\label{2ptSCFT}
\langle0|\hat{\textbf{J}}_s(\lambda_1,\bar{\mu}_1,\psi_1)\hat{\textbf{J}}_s(\lambda_2,\bar{\mu}_2,\psi_2)|0\rangle=\frac{1}{(\lambda_1\cdot\bar{\mu}_2-\lambda_2\cdot\bar{\mu}_1-\psi_1\psi_2)^{2s+2}}.
\end{align}
Implementing the incidence relations \eqref{IncidenceSUSY} in the two point function \eqref{2ptSCFT}, the super-Penrose transform yields
\begin{align}\label{2ptSCFTcon}
\langle0|{\textbf{J}}_s^{a_1\cdots a_{2s}}(x_{1},\theta_1){\textbf{J}}_s^{b_1\cdots b_{2s}}(x_2,\theta_2)|0\rangle&=\int_{\mathbb{RP}^1}D\lambda_1D\lambda_2\prod_{i=1}^{2s}\lambda_1^{a_i}\lambda_2^{b_i}\;\langle0|\hat{\textbf{J}}_s(\lambda_1,\bar{\mu}_1,\psi_1)\hat{\textbf{J}}_s(\bar{\mu}_2.\lambda_2,\psi_2)|0\rangle\bigg\rvert_{\mathfrak{X}}\notag\\
&=\int_{\mathbb{RP}^1}D\lambda_1D\lambda_2\prod_{i=1}^{2s}\lambda_1^{a_i}\lambda_2^{b_i}\;\frac{1}{(\lambda_{1a}\lambda_{2b}(\tilde{X}_{12}^{-})^{ab})^{2s+2}},
\end{align}
where $(\tilde{X}_{12}^{-})^{ab}$ is defined as\footnote{We present several other important position superspace identities in appendix \ref{app:2}.}
\begin{align}
(\tilde{X}_{12}^\pm)^{ab}&=(x_{12})^{ab}+\frac{i}{2}(\theta_{1}^a\theta_{2}^b+\theta_{1}^b\theta_{2}^a)\pm\frac{i}{2}(\theta_{1}^a-\theta_{2}^a)(\theta_{1}^b-\theta_{2}^b).
\end{align}
At this stage, we contract with the polarization spinors and express the integrand in \eqref{2ptSCFTcon} as a Schwinger integral. Then \eqref{2ptSCFTcon} can be recast as 
\begin{align}
&\langle 0|{\mathbf{J}}_s(x_1,\theta_1){\mathbf{J}}_s(x_2,\theta_2)| 0\rangle\notag\\
&=\int_{\mathbb{RP}^1}D\lambda_1D\lambda_2 (\xi_{1}\cdot\lambda_1)^{2s}(\xi_{2}\cdot\lambda_2)^{2s}\int dc_{12}\;\textrm{Sgn}{(c_{12})}c_{12}^{2s+1}\textrm{exp}(-ic_{12}\lambda_{1a}\lambda_{2b}(\tilde{X}_{12}^{-})^{ab})).
\end{align}
This integration can be solved using a derivative operation in the following manner
\begin{align}\label{Susy2mid}
&\langle 0|{\mathbf{J}}_s(x_1,\theta_1){\mathbf{J}}_s(x_2,\theta_2)| 0\rangle\notag\\
&=\int_{\mathbb{RP}^1}D\lambda_1D\lambda_2 (\xi_{1}\cdot\lambda_1)^{2s}\Big(\xi_{2p}((\tilde{X}^{-}_{12})^{pq})^{-1}\frac{\partial}{\partial\lambda_{1}^q}\Big)^{2s}\int dc_{12}\;\textrm{Sgn}{(c_{12})}\;c_{12}\;\textrm{exp}(-ic_{12}\lambda_{1a}\lambda_{2b}(\tilde{X}_{12}^{-})^{ab}),\notag\\
&=\int_{\mathbb{RP}^1}D\lambda_1D\lambda_2  (\xi_{1}\cdot\lambda_1)^{2s}\Big(\xi_{2p}\frac{(\tilde{X}^{+}_{12})^{pq}}{|\tilde{x}_{12}|^2}\frac{\partial}{\partial\lambda_{1}^q}\Big)^{2s}\int dc_{12}\;\textrm{Sgn}{(c_{12})}\;c_{12}\;\textrm{exp}(-ic_{12}\lambda_{1a}\lambda_{2b}(\tilde{X}_{12}^{-})^{ab}),
\end{align}
where, we use the fact that $(\tilde{X}^{-}_{12})_c^a(\tilde{X}^{+}_{12})_b^c=\tilde{x}_{12}^2\delta_b^a$.
Integrating by parts, we have
\begin{align}
&\langle 0|{\mathbf{J}}_s(x_1,\theta_1){\mathbf{J}}_s(x_2,\theta_2)| 0\rangle\notag\\
&=\textrm{TD}+\int_{\mathbb{RP}^1}D\lambda_1D\lambda_2 \int dc_{12}\;\textrm{Sgn}{(c_{12})}\;c_{12}\;\textrm{exp}(-ic_{12}\lambda_{1a}\lambda_{2b}(\tilde{X}_{12}^{-})^{ab})\Big(\frac{\xi_{1m}\xi_{2n}(\tilde{X}^{+}_{12})^{mn}}{|\tilde{x}_{12}|^2}\Big)^{2s},\notag\\
&=\Bigg(\int_{\mathbb{RP}^1}D\lambda_1D\lambda_2 \int dc_{12}\;|c_{12}|\;\textrm{exp}(-ic_{12}\lambda_{1a}\lambda_{2b}(\tilde{X}_{12}^{-})^{ab})\Bigg)\Big(\frac{\xi_{1m}\xi_{2n}(\tilde{X}^{+}_{12})^{mn}}{|\tilde{x}_{12}|^2}\Big)^{2s},
\end{align}
where the total derivative term (TD) is zero. The remaining integrals can be evaluated in the same manner as in non-supersymmetric case \eqref{Mod}
\begin{align}\label{Mod1}
&\int_{\mathbb{RP}^1}D\lambda_1D\lambda_2\int dc_{12}\;\textrm{Sgn}(c_{12})\textrm{exp}(-ic_{12}\lambda_{1a}\lambda_{2b}(\tilde{X}_{12}^{-})^{ab})=\frac{1}{\textrm{Det}|(\tilde{X}_{12}^{-})^{ab}|}=\frac{1}{2\tilde{x}_{12}^2},
\end{align}
Thus, the final result using \eqref{N1} turns out to be the following
\begin{align}\label{2ptSCFTpos}
\langle 0|{\mathbf{J}}_s(x_1,\theta_1){\mathbf{J}}_s(x_2,\theta_2)| 0\rangle=\frac{(\xi_{1m}(\tilde{X}^{-}_{12})^{m}_n\xi_{2}^n)^{2s}}{\tilde{x}_{12}^{4s+2}}=\frac{P^{2s}_{3}}{\tilde{x}^2_{12}},
\end{align}
where $P_3$ is now the supersymmetric invariant building block mentioned in \cite{Nizami:2013tpa}. Thus, the position superspace two point function \eqref{2ptSCFTpos} is obtained from its twistor superspace counterpart \eqref{2ptSCFT} via a super-Penrose transform over. We now move on to three point functions.

\subsection*{Three point functions}
The general form of three point functions is given by \cite{Bala:2025jbh}
\begin{align}\label{ThreePoint}
\langle 0| \mathbf{\hat{J}}_{s_1}(\lambda_1,\bar{\mu}_1,\psi_1)\mathbf{\hat{J}}_{s_2}(\lambda_2,\bar{\mu}_2,\psi_2)\mathbf{\hat{J}}_{s_3}(\lambda_3,\bar{\mu}_3,\psi_3)| 0\rangle=\prod_{\substack{i,j,k=1 \\ i<j\\k\neq i,j}}^3\delta^{[s_i+s_j-s_k]}(\lambda_i\cdot\bar{\mu}_j-\lambda_j\cdot\bar{\mu}_i-\psi_i\psi_j).
\end{align}
Let us begin with the simplest case of three point function of scalar supermultiplets. 
\subsubsection*{$\langle 0|\mathbf{J_0}\mathbf{J_0}\mathbf{J_0}| 0\rangle$}
Consider scalar supermultiplets of scaling dimension $\Delta=1$. The twistor superspace expression for this three point function is given by
\begin{align}\label{OOO}
&\langle 0| \mathbf{\hat{J}_0}(\lambda_1,\bar{\mu}_1,\psi_1)\mathbf{\hat{J}_0}(\lambda_2,\bar{\mu}_2,\psi_2)\mathbf{\hat{J}_0}(\lambda_3,\bar{\mu}_3,\psi_3)| 0\rangle \notag\\
&=\delta(\lambda_1\cdot\bar{\mu}_2-\lambda_2\cdot\bar{\mu}_1-\psi_1\psi_2)\delta(\lambda_2\cdot\bar{\mu}_3-\lambda_3\cdot\bar{\mu}_2-\psi_2\psi_3)\delta(\lambda_3\cdot\bar{\mu}_1-\lambda_1\cdot\bar{\mu}_3-\psi_3\psi_1).
\end{align}
The super-Penrose transform is defined as 
\begin{align}
&\langle 0| \mathbf{J_0}(x_1,\theta_1)\mathbf{J_0}(x_2,\theta_2)\mathbf{J_0}(x_3,\theta_3)| 0\rangle\notag\\
&=\int_{\mathbb{RP}^1}D\lambda_1D\lambda_2D\lambda_3(\xi_1\cdot\lambda_1)^0(\xi_2\cdot\lambda_2)^0(\xi_3\cdot\lambda_3)^0\;\langle 0| \mathbf{\hat{J}_0}(\lambda_1,\bar{\mu}_1,\psi_1)\mathbf{\hat{J}_0}(\lambda_2,\bar{\mu}_2,\psi_2)\mathbf{\hat{J}_0}(\lambda_3,\bar{\mu}_3,\psi_3)| 0\rangle\bigg\rvert_{\mathfrak{X}}\notag\\
&=\int_{\mathbb{RP}^1}D\lambda_1D\lambda_2D\lambda_3\;\delta(\lambda_{1a}\lambda_{2b}(\tilde{X}_{12}^{-})^{ab})\delta(\lambda_{2c}\lambda_{3d}(\tilde{X}_{23}^{-})^{cd})\delta(\lambda_{3e}\lambda_{1f}(\tilde{X}_{31}^{-})^{ef}).
\end{align}
Now we projectively integrate over $\lambda_1$ and $\lambda_2$, which leads to the following expression
\begin{align}\label{OOOSF}
&\langle 0| \mathbf{J_0}(x_1,\theta_1)\mathbf{J_0}(x_2,\theta_2)\mathbf{J_0}(x_3,\theta_3)| 0\rangle=\int_{\mathbb{RP}^1}D\lambda_3\;\delta(\lambda_{3a}(\tilde{X}_{31}^{+})^{a}_b(\tilde{X}_{12}^{+})^{b}_c(\tilde{X}_{23}^{+})^{c}_d\lambda_{3}^d),
\end{align}
where we have used the a simple property of projective integrals
\begin{align}\label{Delta}
\int D\lambda\;f(\lambda)\delta(\lambda\cdot\kappa)=f(\kappa).
\end{align}
Returning to \eqref{OOOSF}, we take note of the fact that the argument in the delta function must be symmetric in the indices `a' and `d' when both of these indices are lowered. Although this was apparent for the non-supersymmetric case \cite{Baumann:2024ttn}, it is not the case for the supersymmetric scenario. This happens to be one of the many crucial nuances associated with the projective integrals in the supersymmetric setting. Thus, we symmetrize the argument as follows 
\begin{align}\label{Sym}
\delta_{Sym}\big(\lambda_{3a}(\tilde{X}_{31}^{+})^{a}_b(\tilde{X}_{12}^{+})^{b}_c(\tilde{X}_{23}^{+})^{c}_d\lambda_{3}^d)&\equiv\frac{1}{2}\;\delta\Bigg(\frac{\lambda_{3a}(\tilde{X}_{31}^{+})^{a}_b(\tilde{X}_{12}^{+})^{b}_c(\tilde{X}_{23}^{+})^{c}_d\lambda_{3}^d+(\lambda_{3a}(\tilde{X}_{31}^{+})^{a}_b(\tilde{X}_{12}^{+})^{b}_c(\tilde{X}_{23}^{+})^{c}_d\lambda_{3}^d)^T}{2}\Bigg)\notag\\
&=\delta\big(\lambda_{3a}((\tilde{X}_{31}^{+})^{a}_b(\tilde{X}_{12}^{+})^{b}_c(\tilde{X}_{23}^{+})^{c}_d+(\tilde{X}_{23}^{-})^{a}_b(\tilde{X}_{12}^{-})^{b}_c(\tilde{X}_{31}^{-})^{c}_d)\lambda_{3}^d\big),
\end{align}
where we have used the fact that $[(\tilde{X}_{ij}^\pm)_a^b]^T=(\tilde{X}_{ij}^\mp)_b^a$. Having symmetrized the argument of the integrand in \eqref{OOOSF}, we notice that the integral is of the following form
\begin{align}\label{Det}
\int D\lambda \, \delta(\lambda \cdot \kappa \cdot \lambda) \propto \frac{1}{\sqrt{|\det \kappa|}}.
\end{align}
Thus, its evaluation results in
\begin{align}\label{OOOF}
\langle 0| \mathbf{J_0}(x_1,\theta_1)\mathbf{J_0}(x_2,\theta_2)\mathbf{J_0}(x_3,\theta_3)| 0\rangle&=\frac{1}{\sqrt{\textrm{Det}((\tilde{X}_{31}^{+})^{a}_b(\tilde{X}_{12}^{+})^{b}_c(\tilde{X}_{23}^{+})^{c}_d+(\tilde{X}_{23}^{-})^{a}_b(\tilde{X}_{12}^{-})^{b}_c(\tilde{X}_{31}^{-})^{c}_d))}}\notag\\
&=\frac{1}{\tilde{x}_{12}\tilde{x}_{23}\tilde{x}_{31}},
\end{align}
which is the three point function for scalar supermultiplets \cite{Nizami:2013tpa}. Thus, we observe that the super-Penrose transform leads to the correct three point functions in position superspace. Let us now move to the cases involving spinning supermultiplets.

\subsubsection*{$\langle 0|\mathbf{J}_{\frac{1}{2}}\mathbf{J}_{\frac{1}{2}}\mathbf{J_0}| 0\rangle$}
We now work out a supercorrelator involving two conserved supercurrents of spin half and one scalar supercurrent of $\Delta=1$. The corresponding three point function in twistor superspace is
\begin{align}\label{JhJhO}
&\langle 0| \mathbf{\hat{J}}_\frac{1}{2}(\lambda_1,\bar{\mu}_1,\psi_1)\mathbf{\hat{J}}_\frac{1}{2}(\lambda_2,\bar{\mu}_2,\psi_2)\mathbf{\hat{J}}_0(\lambda_3,\bar{\mu}_3,\psi_3)| 0\rangle\notag\\
&=\delta^{[1]}(\lambda_1\cdot\bar{\mu}_2-\lambda_2\cdot\bar{\mu}_1-\psi_1\cdot\psi_2)\delta(\lambda_2\cdot\bar{\mu}_3-\lambda_3\cdot\bar{\mu}_2-\psi_2\cdot\psi_3)\delta(\lambda_3\cdot\bar{\mu}_1-\lambda_1\cdot\bar{\mu}_3-\psi_3\cdot\psi_1).
\end{align}
The super-Penrose transform after imposing the incidence relations \eqref{IncidenceSUSY} is
\begin{align}
&\langle 0| \mathbf{J}_\frac{1}{2}(x_1,\theta_1)\mathbf{J}_\frac{1}{2}(x_2,\theta_2)\mathbf{J}_{0}(x_3,\theta_3)| 0\rangle\notag\\
&=\int_{\mathbb{RP}^1}D\lambda_1D\lambda_2D\lambda_3(\xi_1\cdot\lambda_1)(\xi_2\cdot\lambda_2)\;\langle 0| \mathbf{\hat{J}}_\frac{1}{2}(\lambda_1,\bar{\mu}_1,\psi_1)\mathbf{\hat{J}}_\frac{1}{2}(\lambda_2,\bar{\mu}_2,\psi_2)\mathbf{\hat{J}}_0(\lambda_3,\bar{\mu}_3,\psi_3)| 0\rangle\bigg\rvert_\mathfrak{X}\notag\\
&=\int_{\mathbb{RP}^1}D\lambda_1D\lambda_2D\lambda_3(\xi_1\cdot\lambda_1)(\xi_2\cdot\lambda_2)\;\delta^{[1]}(\lambda_{1a}\lambda_{2b}(\tilde{X}_{12}^{-})^{ab})\delta(\lambda_{2c}\lambda_{3d}(\tilde{X}_{23}^{-})^{cd})\delta(\lambda_{3e}\lambda_{1f}(\tilde{X}_{31}^{-})^{ef}).
\end{align}
Performing the projective integrals over $\lambda_2\And\lambda_3$ yields
\begin{align}\label{JhJhO}
&\langle 0| \mathbf{J}_\frac{1}{2}(x_1,\theta_1)\mathbf{J}_\frac{1}{2}(x_2,\theta_2)\mathbf{J}_{0}(x_3,\theta_3)| 0\rangle\notag\\
&=\int_{\mathbb{RP}^1}D\lambda_1\big(\xi_{1m}\lambda_1^m\big)\big(\xi_{2p}(\tilde{X}^{+}_{23})^{p}_{q}(\tilde{X}^{+}_{31})^{q}_{r}\lambda_{1}^r\big)
\delta^{[1]}_{Sym}(\lambda_{1a}(\tilde{X}_{12}^{-})^{a}_b(\tilde{X}_{23}^{-})^{b}_c(\tilde{X}_{31}^{-})^{c}_d\lambda_{1}^d),
\end{align}
where we have symmetrized the argument of the final integral using \eqref{Sym}.

At this stage, we conjure an inverse operator that acts as a derivative on the delta function under the integral, much like in \eqref{Susy2mid} 
\begin{align}\label{Inverse}
\int D\lambda\;\lambda_a\;\delta^{[n]}_{sym}(\lambda\cdot\kappa\cdot\lambda)=\int D\lambda\;\kappa^{-1}_{ab}\frac{\partial}{\partial\lambda_b}\delta^{[n-1]}_{sym}(\lambda\cdot\kappa\cdot\lambda).
\end{align}
For our purposes, we use the fact that the argument `$\kappa$' is symmetric in its indices. Moreover, the square of the argument of the delta function in our case is equal to its determinant. Thus, we define the inverse operator $\kappa^{-1}$ as
\begin{align}\label{InverseOp}
  (\kappa^{-1})_{a}^d =\Bigg(\frac{(\tilde{X}_{12}^{+})^{b}_a(\tilde{X}_{23}^{+})^{c}_b(\tilde{X}_{31}^{+})^{d}_c+(\tilde{X}_{31}^{-})^{b}_a(\tilde{X}_{23}^{-})^{c}_b(\tilde{X}_{12}^{-})^{d}_c}{2\;\tilde{x}_{12}^2\tilde{x}_{23}^2\tilde{x}_{31}^2}\Bigg)
\end{align}
This allows us to insert the inverse operator and re-express \eqref{JhJhO} as a partial integral
\begin{align}
&\langle 0| \mathbf{J}_\frac{1}{2}(x_1,\theta_1)\mathbf{J}_\frac{1}{2}(x_2,\theta_2)\mathbf{J}_{0}(x_3,\theta_3)| 0\rangle\notag\\
&=\int_{\mathbb{RP}^1}D\lambda_1\big(\xi_{1m}\lambda_1^m\big)\big(\xi_{2p}(\tilde{X}^{+}_{23})^{p}_{q}(\tilde{X}^{+}_{31})^{q}_{r}\big)(\kappa^{-1})_{k}^r
\frac{\partial}{\partial \lambda_{1k}}\delta_{Sym}\big(\lambda_{1a}(\tilde{X}_{12}^{+})^{a}_b(\tilde{X}_{23}^{+})^{b}_c(\tilde{X}_{31}^{+})^{c}_d\lambda_{1}^d\big).
\end{align}
An important comment is in order at this stage. The manifest symmetrization of the argument in the delta function requires the addition of the argument and its transpose \eqref{Sym}. This point is reflected even in the construction of the inverse operator \eqref{InverseOp}. Although these two terms turn out to be the same in the non-supersymmetric case, this is not true for the supersymmetric scenario. This subtlety is of utmost importance and plays a crucial role in obtaining the results in position superspace.

Integrating by parts, we have the following expression
\begin{align}
&\langle 0| \mathbf{J}_\frac{1}{2}(x_1,\theta_1)\mathbf{J}_\frac{1}{2}(x_2,\theta_2)\mathbf{J}_{0}(x_3,\theta_3)| 0\rangle\notag\\
&=\int_{\mathbb{RP}^1}D\lambda_1\delta_{Sym}\big(\lambda_{1a}(\tilde{X}_{12}^{+})^{a}_b(\tilde{X}_{23}^{+})^{b}_c(\tilde{X}_{31}^{+})^{c}_d\lambda_{1}^d\big)\big(\xi_{2p}(\tilde{X}^{+}_{23})^{p}_{q}(\tilde{X}^{+}_{31})^{q}_{r}\big)(\kappa^{-1})_{k}^r\frac{\partial}{\partial \lambda_{1k}}\big(\xi_{1m}\lambda_1^m\big),
\end{align}
where the total derivative term is zero. After substituting \eqref{InverseOp} and performing a few simplifications, we perform the final projective integral using \eqref{Det}. This results in
\begin{align}
&\langle 0| \mathbf{J}_\frac{1}{2}(x_1,\theta_1)\mathbf{J}_\frac{1}{2}(x_2,\theta_2)\mathbf{J}_{0}(x_3,\theta_3)| 0\rangle\notag\\
&=\frac{1}{2\tilde{x}_{12}\tilde{x}_{23}\tilde{x}_{31}}\Bigg(\frac{\xi_{1m}(\tilde{X}_{12}^+)^m_n\xi_2^n}{\tilde{x}^2_{12}}\Bigg)+\frac{1}{2\tilde{x}_{12}\tilde{x}_{23}\tilde{x}_{31}}\Bigg(\frac{\xi_{1a}(\tilde{X}_{31}^-)^a_b(\tilde{X}_{23}^-)^b_c(\tilde{X}_{12}^-)^c_d(\tilde{X}_{31}^-)^d_e(\tilde{X}_{23}^-)^e_f\xi_2^f}{\tilde{x}_{12}^2\tilde{x}_{23}^2\tilde{x}_{31}^2}\Bigg).
\end{align}
Although the first term manifestly assumes the building block $P_3$, one can see that the second one is also invariant under supertranslation and superinversion. The key point here is that the relative coefficient between these two terms is fixed due to conservation. Recall that the supercurrents were conserved from the word go in twistor superspace \cite{Bala:2025jbh}, hence this reflects in position superspace as well. We checked numerically at this point that the final result matches with \cite{Nizami:2013tpa}
\begin{align}
\langle 0| \mathbf{J}_\frac{1}{2}(x_1,\theta_1)\mathbf{J}_\frac{1}{2}(x_2,\theta_2)\mathbf{J}_{0}(x_3,\theta_3)| 0\rangle=&\frac{P_3-iR_1R_2}{\tilde{x}_{12}\tilde{x}_{23}\tilde{x}_{31}}.
\end{align}
The other supercorrelators can also be obtained by following a similar algorithm. We present a detailed example for the super-Penrose transform in the appendix \ref{app:4}. Thus, one can obtain the supercorrelators in position superspace from twistor superspace via the super-Penrose transform \eqref{SUSYPenrose}. We now move on to define the super-Witten transform.

\section{Super-Witten transform for $\mathcal{N}=1$}\label{sec:4}
In the section \ref{sec:2}, we saw how the Witten transform \eqref{Witten} maps from twistor space to momentum space via a half-Fourier transform. The super-Witten transform is the supersymmetric generalization of this, which involves a half-Fourier transform over both: the bosonic coordinate `$\bar{\mu}$' along with the fermionic coordinate `$\psi$'. Although this is a straightforward operation in four dimensional case \cite{Witten:2003nn}, it is a little cumbersome for three dimensional supertwistors. This stems from the way in which the Grassmann twistor `$\psi$' is related to the Grassmann spinor `$\theta_a$'. We urge the reader to refer section 3.2 and appendix C of \cite{Jain:2023idr} for a detailed commentary over this matter.

\subsection{Super-Witten transform: Supertwistors $\rightarrow$ momentum superspace}
Consider the case of $\mathcal{N}=1$ supersymmetry in three dimensions. The super-Witten transform maps from the twistor superspace to the spinor-helicity variables in momentum superspace
\begin{align}
(\lambda,\bar{\mu},\psi)\longrightarrow(\lambda,\bar{\lambda},\eta,\bar{\eta}).
\end{align}
For a function $\hat{\mathbf{f}}_s(\lambda,\bar{\mu},\psi)$ of spin `s' in twistor superspace, the super-Witten transform is defined in the following manner\footnote{We describe the algorithm for $\hat{J}_s^+$ here, but one can do the same for $\hat{J}_s^-$ with the variables $\psi,\;\chi,\;$etc. being replaced by $\bar{\psi},\;\bar{\chi},\;$etc. respectively.}
\begin{enumerate}
\item Define $\mathfrak{P}:=\psi=\frac{e^{-\frac{i\pi}{4}}}{2\sqrt{2}}(\chi+\eta)$.
\item Multiply the function by an overall factor of $(\chi-\eta)$.
\item Now, perform the inverse Grassmann twistor transform \cite{Jain:2023idr} from $\chi\rightarrow\bar{\eta}$
\begin{align}\label{GITT1}
\tilde{f}_s(\lambda,\bar{\mu},\eta,\bar{\eta})=\int d\chi\;\textrm{exp}\Big(\frac{\chi\bar{\eta}}{4}\Big)(\chi-\eta)\hat{f}_s\left(\lambda,\bar{\mu},\psi\right)\bigg\rvert_\mathfrak{P}.
\end{align}
\item Augment the Witten transform \eqref{Witten} along with \eqref{GITT1} to complete the transformation
\begin{align}\label{SUSYWitten}
\tilde{f}_s(\lambda,\bar{\lambda},\eta,\bar{\eta})=\int d^2\bar{\mu}\;\textrm{exp}(-i\bar{\lambda}_a\bar{\mu}^a)\int d\chi\;\textrm{exp}\Big(\frac{\chi\bar{\eta}}{4}\Big)(\chi-\eta)\hat{f}_s\left(\lambda,\bar{\mu},\psi\right)\bigg\rvert_\mathfrak{P}.
\end{align}
\end{enumerate}
Equipped with the super-Witten transform \eqref{SUSYWitten}, let us derive the supercorrelators in momentum superspace.
\subsection{Examples}
We begin with the simple case two point functions.
\subsection*{Two Point Function}
Recall the two point function for conserved supercurrents of spin `s' in twistor superspace \eqref{2ptSCFT}. It takes the following form when expressed as a Schwinger integral
\begin{align}\label{2ptSCFT1}
 \langle0|\hat{\textbf{J}}_s(\lambda_1,\bar{\mu}_1,\psi_1)\hat{\textbf{J}}_s(\lambda_2,\bar{\mu}_2,\psi_2)|0\rangle=&\int dc_{12}\;c_{12}^{2s+1}\textrm{Sgn}(c_{12})\;\textrm{exp}(ic_{12}(\lambda_1\cdot\bar{\mu}_2-\lambda_2\cdot\bar{\mu}_1-\psi_1\psi_2))\notag\\
 =&\int dc_{12}\;c_{12}^{2s+1}\textrm{Sgn}(c_{12})\left(1-ic_{12}\psi_1\psi_2\right)\;\textrm{exp}(ic_{12}(\lambda_1\cdot\bar{\mu}_2-\lambda_2\cdot\bar{\mu}_1)).
\end{align}
The momentum space supercorrelator for \eqref{2ptSCFT1} can be obtained by performing a super-Witten transform \eqref{SUSYWitten} in the following manner
\begin{align}
&\langle0|\tilde{\textbf{J}}_s(\lambda_1,\bar{\lambda}_1,\eta_1,\bar{\eta}_1)\tilde{\textbf{J}}_s(\lambda_2,\bar{\lambda}_2,\eta_2,\bar{\eta}_2)|0\rangle
\notag\\&=\prod_{i=1}^2\int d^2\bar{\mu}_i\;\textrm{exp}(-i\bar{\lambda}_{ia}\bar{\mu}_i^a)\int d\chi_i\;\textrm{exp}\Big(\frac{\chi_i\bar{\eta}_i}{4}\Big)(\chi_i-\eta_i)\langle0|\hat{\textbf{J}}_s(\lambda_1,\bar{\mu}_1,\psi_1)\hat{\textbf{J}}_s(\lambda_2,\bar{\mu}_2,\psi_2)|0\rangle\bigg\rvert_{\mathfrak{P}}.
\end{align}
We begin by performing the familiar half-Fourier transform over the bosonic coordinates, which is the Witten transform \eqref{Witten}. This leads to the expression\footnote{We suppress the momentum-conserving delta function to avoid clutter.}
\begin{align}
&\langle0|\tilde{\textbf{J}}_s(\lambda_1,\bar{\lambda}_1,\eta_1,\bar{\eta}_1)\tilde{\textbf{J}}_s(\lambda_2,\bar{\lambda}_2,\eta_2,\bar{\eta}_2)|0\rangle\notag\\
&=\int d\chi_1d\chi_2\;\textrm{exp}\Big(\frac{\chi_1\bar{\eta}_1}{4}\Big)\textrm{exp}\Big(\frac{\chi_2\bar{\eta}_2}{4}\Big)(\chi_1-\eta_1)(\chi_2-\eta_2)\frac{\langle\bar{1}\bar{2}\rangle^{2s}}{(2p_1)^{2s-1}}\left(1-\frac{(\chi_1+\eta_1)(\chi_2+\eta_2)}{8}\frac{\langle\bar{1}\bar{2}\rangle}{2p_1}\right).
\end{align}
Performing the Grassmann integrals over $\chi_i$ results in
\begin{align}
&\langle0|\tilde{\textbf{J}}_s(\lambda_1,\bar{\lambda}_1,\eta_1,\bar{\eta}_1)\tilde{\textbf{J}}_s(\lambda_2,\bar{\lambda}_2,\eta_2,\bar{\eta}_2)|0=\frac{\langle\bar{1}\bar{2}\rangle^{2s}}{(2p_1)^{2s-1}}\textrm{exp}\left(\frac{\eta_1\bar{\eta}_1}{4}+\frac{\eta_2\bar{\eta}_2}{4}\right)\left(1-\eta_1\eta_2\frac{\langle\bar{1}\bar{2}\rangle}{4p_1}\right).
\end{align}
This is precisely the momentum space supercorrelator in spinor-helicity variables, as noted in \cite{Jain:2023idr}. We now perform the same analysis for three point functions.

\subsection*{Three Point Function}  
Recall the three point supercorrelator of conserved supercurrents \eqref{ThreePoint} is given by 
\begin{align}
&\langle 0| \mathbf{\hat{J}}_{s_1}(\lambda_1,\bar{\mu}_1,\psi_1)\mathbf{\hat{J}}_{s_2}(\lambda_2,\bar{\mu}_2,\psi_2)\mathbf{\hat{J}}_{s_3}(\lambda_3,\bar{\mu}_3,\psi_3)| 0\rangle\notag\\
&=\int dc_{12}dc_{23}dc_{31}\;c_{12}^{s_1+s_2-s_3}c_{23}^{s_2+s_3-s_1}c_{31}^{s_3+s_1-s_2}\;\textrm{exp}(ic_{12}(\lambda_1\cdot\bar{\mu}_2-\lambda_2\cdot\bar{\mu}_1-\psi_1\psi_2)+\textrm{cyclic})\notag\\
&=\int dc_{12}dc_{23}dc_{31}\;c_{12}^{s_1+s_2-s_3}c_{23}^{s_2+s_3-s_1}c_{31}^{s_3+s_1-s_2}(1-ic_{12}\psi_1\psi_2)(1-ic_{23}\psi_2\psi_3)(1-ic_{31}\psi_3\psi_1)\notag\\
&\quad\quad\textrm{exp}(i(c_{12}(\lambda_1\cdot\bar{\mu}_2-\lambda_2\cdot\bar{\mu}_1)+c_{23}(\lambda_2\cdot\bar{\mu}_3-\lambda_3\cdot\bar{\mu}_2)+c_{31}(\lambda_3\cdot\bar{\mu}_1-\lambda_1\cdot\bar{\mu}_3)).
\end{align}
The momentum superspace supercorrelator can be obtained by performing a super-Witten transform \eqref{SUSYWitten} in the following manner
\begin{align}\label{3ptSW}
\notag&\langle0|\tilde{\textbf{J}}_{s_1}(\lambda_1,\bar{\lambda}_1,\eta_1,\bar{\eta}_1)\tilde{\textbf{J}}_{s_2}(\lambda_2,\bar{\lambda}_2,\eta_2,\bar{\eta}_2)\tilde{\textbf{J}}_{s_3}(\lambda_3,\bar{\lambda}_3,\eta_3,\bar{\eta}_3)|0\rangle\notag\\
&=\prod_{i=1}^3\int d^2\bar{\mu}_i\;\textrm{exp}(-i\bar{\lambda}_{ia}\bar{\mu}_i^a)\int d\chi_i\;\textrm{exp}\Big(\frac{\chi_i\bar{\eta}_i}{4}\Big)(\chi_i-\eta_i)\langle 0| \mathbf{\hat{J}}_{s_1}(\lambda_1,\bar{\mu}_1,\psi_1)\mathbf{\hat{J}}_{s_2}(\lambda_2,\bar{\mu}_2,\psi_2)\mathbf{\hat{J}}_{s_3}(\lambda_3,\bar{\mu}_3,\psi_3)| 0\rangle\bigg\rvert_{\mathfrak{P}}.
\end{align}
Performing the Witten transform \eqref{Witten} over the bosonic coordinates results in the supercorrelator in the spinor-helicity variables. The details of this computation can be found in appendix \ref{app:5}. This integration leads to the expression
\begin{align}
&\langle0|\tilde{\textbf{J}}_{s_1}(\lambda_1,\bar{\lambda}_1,\eta_1,\bar{\eta}_1)\tilde{\textbf{J}}_{s_2}(\lambda_2,\bar{\lambda}_2,\eta_2,\bar{\eta}_2)\tilde{\textbf{J}}_{s_3}(\lambda_3,\bar{\lambda}_3,\eta_3,\bar{\eta}_3)|0\rangle\notag\\
&=\prod_{m=1}^3\int d\chi_m\;\textrm{exp}\Big(\frac{\chi_m\bar{\eta}_m}{4}\Big)(\chi_m-\eta_m)\prod_{\substack{i,j,k=1 \\ i<j\\k\neq i,j}}^3\left(1-\frac{(\chi_i+\eta_i)(\chi_j+\eta_j)}{8}\frac{\langle\bar{i}\bar{j}\rangle}{E}\right)\left(\frac{\langle\bar{i}\bar{j}\rangle}{E}\right)^{s_i+s_j-s_k},
\end{align}
where $E$ is the total energy i.e. $E=p_1+p_2+p_3$. Finally, performing the Grassmann integrals over $\chi_i$ results in
\begin{align}\label{JJJ}
&\langle0|\tilde{\textbf{J}}_{s_1}(\lambda_1,\bar{\lambda}_1,\eta_1,\bar{\eta}_1)\tilde{\textbf{J}}_{s_2}(\lambda_2,\bar{\lambda}_2,\eta_2,\bar{\eta}_2)\tilde{\textbf{J}}_{s_3}(\lambda_3,\bar{\lambda}_3,\eta_3,\bar{\eta}_3)|0\rangle\notag\\&=\frac{\langle\bar{1}\bar{2}\rangle^{s_1+s_2-s_3}\langle\bar{2}\bar{3}\rangle^{s_2+s_3-s_1}\langle\bar{3}\bar{1}\rangle^{s_3+s_1-s_2}}{E^{s_1+s_2+s_3}}
\textrm{exp}\left(\frac{\eta_1\bar{\eta}_1+\eta_2\bar{\eta}_2+\eta_3\bar{\eta}_3}{4}\right)\left(1-\eta_1\eta_2\frac{\langle\bar{1}\bar{2}\rangle}{2E}-\eta_2\eta_3\frac{\langle\bar{2}\bar{3}\rangle}{2E}-\eta_3\eta_1\frac{\langle\bar{3}\bar{1}\rangle}{2E}\right).
\end{align}
This matches the result presented in \cite{Jain:2023idr}. While here we focused on the (+++) helicity correlators to illustrate the algorithm, a similar analysis can be done to obtain expressions for other helicity configurations, too. Thus, one can obtain the supercorrelators in momentum superspace from twistor superspace via the super-Witten transform \eqref{SUSYWitten}. We will now define these transformations for supercorrelators with extended supersymmetries.

\section{The case of extended supersymmetries: $\mathcal{N}=\{2,3,4\}$}\label{sec:5}

Extended supersymmetries posit an important arena of study, namely with $\mathcal{N}=6$ ABJM theories in three dimensions. Although manifest superspace techniques are not possible for such a high extent of supersymmetry, they form an important stepping stone in this direction. The traditional position superspace formalism \cite{Buchbinder:2015qsa,Buchbinder:2015wia,Kuzenko:2016cmf,Jain:2022izp} is crucial to that end, but they become progressively challenging with a large number of tensor structures to deal with. Recently, it was shown in \cite{Bala:2025jbh} that a supertwistor formalism is more amicable for studying supercorrelators up to $\mathcal{N}=4$ supersymmetry in three dimensions. Using twistor superspace as the starting point, we develop the super-Penrose and super-Witten transforms for supercorrelators enjoying extended supersymmetry, starting with the former.

\subsection{Super-Penrose transform}
The super-Penrose transform for extended supersymmetry follows the same suit as in the $\mathcal{N}=1$ case \eqref{SUSYPenrose}. It is a projective integral over $\mathbb{RP}^1$ subject to the appropriate incidence relations for the bosonic and fermionic coordinates $\bar{\mu}\;\textrm{and}\;\psi_A$, respectively
\begin{align}\label{IncidenceN}
\mathfrak{X}_A:=\qquad\Bigg(\bar{\mu}^a=x^{a}_b\lambda^b-\frac{i}{4}\delta^{AB}\theta_{kA}\theta^k_B\lambda^a,\qquad
\psi_A=e^{-\frac{i\pi}{4}}\theta_A^a\lambda_a\Bigg),    
\end{align}
where `$A$' is the R-symmetry index.
Consider a function $\hat{\mathbf{f}}_s(\lambda,\bar{\mu},\psi_A)$ of spin `s'. The super-Penrose transform is defined as
\begin{align}\label{PenroseN}
f_s^{a_1\cdots a_{2s}}(x,\theta_A)=\int_{\mathbb{RP}^1}D\lambda \prod_{i=1}^{2s}\lambda_i^{a_i}\;\hat{f}_s(\lambda,\bar{\mu},\psi_A)\bigg\rvert_{\mathfrak{X}_A}.
\end{align}
Equipped with the super-Penrose transform \eqref{PenroseN}, let us now obtain supercorrelators in position superspace.
\subsubsection{Examples}
Let us start with the simple case involving scalar multiplets.
\subsubsection*{$\langle 0|\mathbf{J_0}\mathbf{J_0}\mathbf{J_0}| 0\rangle$}
Consider the three point function of scalar superfields with scaling dimension $\Delta=1$ in the twistor superspace
\begin{align}\label{OOON}
&\langle 0| \mathbf{\hat{J}}_0(\lambda_1,\bar{\mu}_1,\psi_1^A)\mathbf{\hat{J}}_0(\lambda_2,\bar{\mu}_2,\psi_2^B)\mathbf{\hat{J}}_0(\lambda_3,\bar{\mu}_3,\psi_3^C)| 0\rangle \notag\\
&=\delta(\lambda_1\cdot\bar{\mu}_2-\lambda_2\cdot\bar{\mu}_1-\delta_{AB}\psi_1^A\psi_2^B)\delta(\lambda_2\cdot\bar{\mu}_3-\lambda_3\cdot\bar{\mu}_2-\delta_{BC}\psi_2^B\psi_3^C)\delta(\lambda_3\cdot\bar{\mu}_1-\lambda_1\cdot\bar{\mu}_3-\delta_{CA}\psi_3^C\psi_1^A).
\end{align}
Imposing the incidence relations \eqref{IncidenceN} results in
\begin{align}\label{OOONMid}
&\langle 0| \hat{\mathbf{J}}_0(\lambda_1,x_1,\theta_1^A)\hat{\mathbf{J}}_0(\lambda_2,x_2,\theta_2^B))\hat{\mathbf{J}}_0(\lambda_3,x_3,\theta_3^C))| 0\rangle=\delta(\lambda_{1a}\lambda_{2b}(\tilde{X}_{12}^{-})^{ab})\delta(\lambda_{2c}\lambda_{3d}(\tilde{X}_{23}^{-})^{cd})\delta(\lambda_{3e}\lambda_{1f}(\tilde{X}_{31}^{-})^{ef}),
\end{align}
where \begin{align}
(\tilde{X}_{12}^\pm)^{ab}=(x_{12})^{ab}+\frac{i}{2}\delta^{AB}(\theta_{1A}^a\theta_{2B}^{b}+\theta_{1A}^{b}\theta_{2B}^{a})\pm\frac{i}{2}\delta^{AB}(\theta_{1,A}^{a}-\theta_{2,A}^{a})(\theta_{1,B}^{b}-\theta_{2,B}^{b}).
\end{align}
The super-Penrose transform for \eqref{OOONMid} is as follows
\begin{align}
&\langle 0| {\mathbf{J}}_0(x_1,\theta_1^A)){\mathbf{J}}_0(x_2,\theta_2^B)){\mathbf{J}}_0(x_3,\theta_3^C))| 0\rangle\notag\\
&=\int_{\mathbb{RP}^1}D\lambda_1D\lambda_2D\lambda_3\;\delta(\lambda_{1a}\lambda_{2b}(\tilde{X}_{12}^{-})^{ab})\delta(\lambda_{2c}\lambda_{3d}(\tilde{X}_{23}^{-})^{cd})\delta(\lambda_{3e}\lambda_{1f}(\tilde{X}_{31}^{-})^{ef})\notag\\
&=\int_{\mathbb{RP}^1}D\lambda_3\;\delta_{Sym}(\lambda_{3a}(\tilde{X}_{31}^{+})^{a}_b(\tilde{X}_{12}^{+})^{b}_c(\tilde{X}_{23}^{+})^{c}_d\lambda_{3}^d),
\end{align}
where we have performed the projective integrals over $\lambda_1$ and $\lambda_2$, and symmetrized the argument of the delta function using \eqref{Sym}. The final projective integral results in the following determinant using \eqref{Det}
\begin{align}\label{OOONDet}
&\langle 0|{\mathbf{J}}_0(x_1,\theta_1^A)){\mathbf{J}}_0(x_2,\theta_2^B)){\mathbf{J}}_0(x_3,\theta_3^C))| 0\rangle=\frac{1}{\sqrt{\textrm{Det}((\tilde{X}_{31}^{+})^{a}_b(\tilde{X}_{12}^{+})^{b}_c(\tilde{X}_{23}^{+})^{c}_d+(\tilde{X}_{23}^{-})^{a}_b(\tilde{X}_{12}^{-})^{b}_c(\tilde{X}_{31}^{-})^{c}_d)}}.
\end{align}
The determinant in \eqref{OOONDet}, due to its enhanced supersymmetry, is essentially different from that in \eqref{OOOF}. Upon comparing the result for the $\mathcal{N}=2$ case with \cite{Jain:2022izp} numerically, we find a match with their answer. The supercorrelator for the scalar superfield is
\begin{align}\label{OOONFinal}
&\langle 0| \tilde{\mathbf{J}}_0(x_1,\theta_1^A))\tilde{\mathbf{J}}_0(x_2,\theta_2^B))\tilde{\mathbf{J}}_0(x_3,\theta_3^C))| 0\rangle=\frac{1}{\bar{x}_{12}\bar{x}_{23}\bar{x}_{31}}\Big(1+\frac{R^{'}}{16}\Big).
\end{align}
Observe that \eqref{OOONFinal} has an extra structure $(R')$ due to extended supersymmetry, compared to its analog \eqref{OOOF} in the $\mathcal{N}=1$ case. This is the first of many R-symmetry invariant structures that arise due to extended supersymmetry \cite{Jain:2022izp}. We observe that the super-Penrose transform \eqref{PenroseN}, which is a simple generalization of its $\mathcal{N}=1$ counterpart \eqref{SUSYPenrose}, naturally accommodates these non-trivialities associated with higher supersymmetric cases. Moreover, the above result \eqref{OOONDet} will be true for $\mathcal{N}=3\;\textrm{and}\;4$ as well, provided that the scalar superfields are not charged under R-symmetry.

\subsubsection*{$\langle 0|\mathbf{J}\mathbf{J_0}\mathbf{J_0}| 0\rangle$}
We now move on to the supercorrelator of a conserved spin-1 supercurrent along with two scalar superfields, where all of them are R-symmetry singlets. Its twistor superspace expression is as follows
\begin{align}\label{JOON1}
&\langle 0| \mathbf{\hat{J}}(\lambda_1,\bar{\mu}_1,\psi_1^A)\mathbf{\hat{J}}_0(\lambda_2,\bar{\mu}_2,\psi_2^B)\mathbf{\hat{J}}_0(\lambda_3,\bar{\mu}_3,\psi_3^C)| 0\rangle \notag\\
&=\delta^{[1]}(\lambda_1\cdot\bar{\mu}_2-\lambda_2\cdot\bar{\mu}_1-\delta_{AB}\psi_1^A\psi_2^B)\delta^{[-1]}(\lambda_2\cdot\bar{\mu}_3-\lambda_3\cdot\bar{\mu}_2-\delta_{BC}\psi_2^B\psi_3^C)\delta^{[1]}(\lambda_3\cdot\bar{\mu}_1-\lambda_1\cdot\bar{\mu}_3-\delta_{CA}\psi_3^C\psi_1^A).
\end{align}
The super-Penrose transform \eqref{PenroseN} for the supercorrelator \eqref{JOON1} is as follows
\begin{align}\label{JOON2}
&\langle 0| {\mathbf{J}}(x_1,\theta_1^A){\mathbf{J}}_0(x_2,\theta_2^B){\mathbf{J}}_0(x_3,\theta_3^C)| 0\rangle\notag\\
&=\int_{\mathbb{RP}^1}D\lambda_1D\lambda_2D\lambda_3\;(\xi_1\cdot\lambda_1)^2\;\langle 0| \hat{\mathbf{J}}(\lambda_1,x_1,\theta_1^A)\hat{\mathbf{J}}_0(\lambda_2,x_2,\theta_2^B)\hat{\mathbf{J}}_0(\lambda_3,x_3,\theta_3^C)| 0\rangle\bigg\rvert_\mathfrak{X}.
\end{align}
The prescription for evaluating these integrals remains fairly similar to the $\mathcal{N}=1$ scenario presented in appendix \ref{app:4}. We sketch out some key steps of the derivation here
\begin{enumerate}
\item Impose the incidence relations \eqref{IncidenceN} and express the correlator as a Schwinger integral 
\begin{align}\label{JOON3}
\int_{\mathbb{RP}^1}D\lambda_1D\lambda_2D\lambda_3\;(\xi_1\cdot\lambda_1)^2\int dc_{12}dc_{23}dc_{31}\frac{ic_{12}c_{31}}{c_{23}}\;\textrm{exp}(-ic_{12}\lambda_{1a}\lambda_{2b}(\tilde{X}_{12}^{-})^{ab}+\textrm{cyclic}).
\end{align}
\item Re-express the integrand in \eqref{JOON3} by substituting derivative operator
\begin{align}
\frac{ic_{12}c_{31}}{c_{23}}(\xi_1\cdot\lambda_1)^2\;\textrm{exp}(-ic_{12}\lambda_{1a}\lambda_{2b}(\tilde{X}_{12}^{-})^{ab})\rightarrow&\frac{c_{31}}{c_{23}}(\xi_1\cdot\lambda_1)\Big(\xi_{1m}\frac{(\tilde{X}^{-}_{12})^{mn}}{\bar{x}_{12}^2}\frac{\partial}{\partial \lambda_2^n}\textrm{exp}(-ic_{12}\lambda_{1a}\lambda_{2b}(\tilde{X}_{12}^{-})^{ab})\Big)
\end{align}
and perform partial integrations in the Schwinger parameters
\begin{align}
&(\xi_1\cdot\lambda_1)\Big(\xi_{1m}\frac{(\tilde{X}^{+}_{12})^{m}_{n}(\tilde{X}^{+}_{23})^{n}_{p}}{\bar{x}_{12}^2}\lambda_{3}^p\Big)\delta(\lambda_{1a}\lambda_{2b}(\tilde{X}_{12}^{-})^{ab})\delta(\lambda_{2c}\lambda_{3d}(\tilde{X}_{23}^{-})^{cd})\delta^{[1]}(\lambda_{3e}\lambda_{1f}(\tilde{X}_{31}^{-})^{ef}).
\end{align}
\item Projectively integrate over $\lambda_1$ and $\lambda_2$, and then symmetrize the argument of the remaining delta function
\begin{align}\label{JOONLast}
\int_{\mathbb{RP}^1}D\lambda_3\big(\xi_{1p}(\tilde{X}^{+}_{12})^{p}_{q}(\tilde{X}^{+}_{23})^{q}_{r}\lambda_3^r\big)\Big(\xi_{1m}\frac{(\tilde{X}^{+}_{12})^{m}_{n}(\tilde{X}^{+}_{23})^{n}_{o}}{\bar{x}_{12}^2}\lambda_{3}^o\Big)\delta^{[1]}_{Sym}(\lambda_{3a}(\tilde{X}_{31}^{+})^{a}_b(\tilde{X}_{12}^{+})^{b}_c(\tilde{X}_{23}^{+})^{c}_d\lambda_{3}^d).
\end{align}
\item Insert the appropriate inverse operator \eqref{Inverse}
\begin{align}
&\big(\xi_{1p}(\tilde{X}^{+}_{12})^{p}_{q}(\tilde{X}^{+}_{23})^{q}_{r}\lambda_3^r\big)\delta^{[1]}_{Sym}(\lambda_{3a}(\tilde{X}_{31}^{+})^{a}_b(\tilde{X}_{12}^{+})^{b}_c(\tilde{X}_{23}^{+})^{c}_d\lambda_{3}^d)\notag\\
\longrightarrow&\big(\xi_{1p}(\tilde{X}^{+}_{12})^{p}_{q}(\tilde{X}^{+}_{23})^{q}_{r}\big)(\kappa^{-1})_k^r\frac{\partial}{\partial \lambda_{3k}}\delta_{Sym}\big(\lambda_{3a}(\tilde{X}_{31}^{+})^{a}_b(\tilde{X}_{12}^{+})^{b}_c(\tilde{X}_{23}^{+})^{c}_d\lambda_{3}^d\big),
\end{align}
where
\begin{align}
(\kappa^{-1})_k^r= \Bigg(\frac{(\tilde{X}_{31}^{+})^{r}_i(\tilde{X}_{12}^{+})^{i}_j(\tilde{X}_{23}^{+})^{j}_k+(\tilde{X}_{23}^{-})^{r}_i(\tilde{X}_{12}^{-})^{i}_j(\tilde{X}_{31}^{-})^{j}_k}{{\textrm{Det}((\tilde{X}_{31}^{+})^{a}_b(\tilde{X}_{12}^{+})^{b}_c(\tilde{X}_{23}^{+})^{c}_d+(\tilde{X}_{23}^{-})^{a}_b(\tilde{X}_{12}^{-})^{b}_c(\tilde{X}_{31}^{-})^{c}_d)}}\Bigg),
\end{align}
and perform partial integrations.
\item Evaluate the final integral using \eqref{Det}. This leads to the position superspace result
\begin{align}\label{JOONFinal}
\langle 0| {\mathbf{J}}(x_1,\theta_1^A){\mathbf{J}}_0(x_2,\theta_2^B){\mathbf{J}}_0(x_3,\theta_3^C)| 0\rangle&=\frac{{Q}_1}{\textrm{Det}((\tilde{X}_{31}^{+})^{a}_b(\tilde{X}_{12}^{+})^{b}_c(\tilde{X}_{23}^{+})^{c}_d+(\tilde{X}_{23}^{-})^{a}_b(\tilde{X}_{12}^{-})^{b}_c(\tilde{X}_{31}^{-})^{c}_d)^{\large\frac{3}{2}}}\normalsize,
\end{align}
\end{enumerate}
which matches the $\mathcal{N}=2$ answer presented in \cite{Jain:2022izp}.
Hence, one can obtain the position superspace correlation functions from their twistor superspace counterparts using the super-Penrose transform for extended supersymmetries, too. Let us now establish the super-Witten transform for extended supersymmetries.

\subsection{Super-Witten transform}
The super-Witten transform for extended supersymmetry has a similar algorithm as in the $\mathcal{N}=1$ case \eqref{SUSYWitten}. Consider a function $\hat{\mathbf{f}}_s(\lambda,\bar{\mu},\psi_A)$ of spin `s'. The super-Witten transform is defined as follows
\begin{enumerate}
\item Define $\mathfrak{P}^A:=\psi^A=\frac{e^{-\frac{i\pi}{4}}}{2\sqrt{2}}(\chi^A+\eta^A)$.
\item Multiply by an overall factor of $(\chi^B-\eta^B)$.
\item Now perform the inverse Grassmann twistor transform from $\chi^C\rightarrow\bar{\eta}^C$ as follows
\begin{align}\label{GITT}
\tilde{f}_s(\lambda,\bar{\mu},\eta^A,\bar{\eta}^A)=\int d\chi^B\;\textrm{exp}\Big(\frac{\chi^C\bar{\eta}^C}{4}\Big)(\chi^B-\eta^B)\hat{f}_s\left(\lambda,\bar{\mu},\psi^A\right)\bigg\rvert_{\mathfrak{P}^A}.
\end{align}
\item Augment the Witten transform \eqref{Witten} with \eqref{GITT} to complete the transformation
\begin{align}\label{WittenN}
\tilde{f}_s(\lambda,\bar{\lambda},\eta^A,\bar{\eta}^A)=\int d^2\bar{\mu}\;\textrm{exp}(-i\bar{\lambda}_a\bar{\mu}^a)\int d\chi^B\;\textrm{exp}\Big(\frac{\chi^C\bar{\eta}^C}{4}\Big)(\chi^B-\eta^B)\hat{f}_s\left(\lambda,\bar{\mu},\psi^A\right)\bigg\rvert_{\mathfrak{P}^A}.
\end{align}
\end{enumerate}
Let us now perform this operation to obtain supercorrelators in momentum superspace.
\subsubsection{Examples}
Consider the three point function for conserved supercurrents. The super-Witten transform \eqref{WittenN} for this supercorrelator is as follows
\begin{align}
&\langle0|\tilde{\textbf{J}}_{s_1}(\lambda_1,\bar{\lambda}_1,\eta_1^A,\bar{\eta}_1^A)\tilde{\textbf{J}}_{s_2}(\lambda_2,\bar{\lambda}_2,\eta_2^B,\bar{\eta}_2^B)\tilde{\textbf{J}}_{s_3}(\lambda_3,\bar{\lambda}_3,\eta_3^C,\bar{\eta}_3^C)|0\rangle=\prod_{i=1}^3\int d^2\bar{\mu}_i\;\textrm{exp}(-i\bar{\lambda}_{ia}\bar{\mu}_i^a)\notag\\
&\int d\chi_i^M\;\textrm{exp}\Big(\frac{\chi_i^N\bar{\eta}_i^N}{4}\Big)(\chi_i^M-\eta_i^M)\langle 0| \mathbf{\hat{J}}_{s_1}(\lambda_1,\bar{\mu}_1,\psi_1^A)\mathbf{\hat{J}}_{s_2}(\lambda_2,\bar{\mu}_2,\psi_2^B)\mathbf{\hat{J}}_{s_3}(\lambda_3,\bar{\mu}_3,\psi_3^C)| 0\rangle\bigg\rvert_{\mathfrak{P}^J}.
\end{align}
Performing the Witten transform \eqref{Witten} over the bosonic coordinates results in the momentum space component correlators, yielding the following expression\footnote{We suppress the sum over R-symmetry indices with `$\cdot$' to avoid clutter.}
\begin{align}
&\langle0|\tilde{\textbf{J}}_{s_1}(\lambda_1,\bar{\lambda}_1,\eta_1^A,\bar{\eta}_1^A)\tilde{\textbf{J}}_{s_2}(\lambda_2,\bar{\lambda}_2,\eta_2^B,\bar{\eta}_2^B)\tilde{\textbf{J}}_{s_3}(\lambda_3,\bar{\lambda}_3,\eta_3^C,\bar{\eta}_3^C)|0\rangle\notag\\
&={\prod_{m=1}^3\int d\chi_m^P\;\textrm{exp}\Big(\frac{\chi_m\cdot\bar{\eta}_m}{4}\Big)(\chi_m^P-\eta_m^P)\prod_{\substack{i,j,k=1 \\ i<j\\k\neq i,j}}^3\left(1-\frac{(\chi_i+\eta_i)\cdot(\chi_j+\eta_j)}{8}\frac{\langle\bar{i}\bar{j}\rangle}{E}\right)\left(\frac{\langle\bar{i}\bar{j}\rangle}{E}\right)^{s_i+s_j-s_k},}
\end{align}
We finally perform the Grassmann integrals over $\chi_i^A$ using the identity
\begin{align}
 \int d\chi_i^A(\chi_i^A-\eta_i^A)f(\chi_i^M)=\int d\chi_i^A\delta^{\mathcal{N}}(\chi_i^A-\eta_i^A)f(\chi_i^M)=f(\eta_i^M),  
\end{align}
which results in the expression
\begin{align}
&{\langle0|\tilde{\textbf{J}}_{s_1}(\lambda_1,\bar{\lambda}_1,\eta_1^A,\bar{\eta}_1^A)\tilde{\textbf{J}}_{s_2}(\lambda_2,\bar{\lambda}_2,\eta_2^B,\bar{\eta}_2^B)\tilde{\textbf{J}}_{s_3}(\lambda_3,\bar{\lambda}_3,\eta_3^C,\bar{\eta}_3^C)|0\rangle}
\notag\\&{=\frac{\langle\bar{1}\bar{2}\rangle^{s_1+s_2-s_3}\langle\bar{2}\bar{3}\rangle^{s_2+s_3-s_1}\langle\bar{3}\bar{1}\rangle^{s_3+s_1-s_2}}{E^{s_1+s_2+s_3}}\textrm{exp}\Big(\sum_{k=1}^3\frac{\eta_k\cdot\bar{\eta}_k}{4}\Big)\left(1-\sum_{\substack{i,j=1 \\ i<j}}^3\eta_i\cdot\eta_j\frac{\langle\bar{i}\bar{j}\rangle}{2E}\right).}
\end{align}
Notice that the result turns out to be a very straightforward extension of the $\mathcal{N}=1$ result \eqref{JJJ}. One can check that this satisfies all the superconformal Ward identities when solved in the momentum superspace.

\section{Discussion and future direction}\label{Discussion}
In this paper, we have developed supersymmetric versions of the Penrose and Witten transforms for three dimensional superconformal field theories with varied amounts of supersymmetry. The former allows to obtain expressions in the position superspace from their twistor superspace counterparts, while the latter leads to the corresponding momentum superspace versions. We demonstrate the utility of these transformations by deriving two and three point supercorrelators directly from twistor superspace results. Furthermore, we extend these constructions to theories with enhanced supersymmetry. Although the position superspace results have a large number of invariant structures to deal with in extended supersymmetries, deriving the result via the super-Penrose transform circumvents this difficulty.
There are several interesting avenues for further exploration, some of which are outlined below.

\subsection*{Non-conserved and R-charged supercurrents}
In this work, we investigated supercorrelators of conserved supercurrents that are R-symmetry singlets. It would be interesting from a technical standpoint to extend our analysis to more generic supermultiplets. This shall greatly aid in studying BPS states in CFT. This will amount to studying the supersymmetric version of \cite{Bala:2025qxr} in the twistor superspace.

\subsection*{Higher point functions}
Another important direction is the construction of higher-point functions in CFTs. In four dimensions, spinor-helicity methods \cite{Britto:2005fq} and twistor techniques \cite{Mason:2009sa} have been instrumental in recursively computing higher point tree-level amplitudes. Extending such recursion techniques to three dimensional CFTs will be crucial in this endeavor.

\subsection*{Supersymmetric Chern-Simons matter theories}
Supersymmetric Chern-Simons matter theories provide a quintessential example where higher spin symmetry is slightly broken. Among these, the $\mathcal{N}=6$ ABJM theory is of paramount importance as it is the maximally supersymmetric version of it. Although a fully manifest study of this theory is not possible, it is desirable to analyze its supercorrelators. One promising approach to tackle this problem is through the harmonic superspace formalism. We hope to revisit this issue and explore it further in the near future.

\acknowledgments
The author gratefully acknowledges Sachin Jain for valuable discussions and comments during the course of this project. The author also thanks Aswini Bala, K.S. Dhruva, and Nipun Bhave for their helpful discussions.

\appendix

\section{Notations and conventions}\label{app:1}
In this appendix, we list the notation and conventions followed in this work. We work with Minkowski space Wightman functions denoted by $\langle0|\cdots|0\rangle$, which is the vacuum expectation value of a particular ordered string of operators.

The non-supersymmetric/component currents are presented in italics (e.g. `$J_s$'), whereas the supercurrents are always presented in bold typeface (e.g. `$\mathbf{J}_s$'). Furthermore, we use `$\hat{\mathbf{J}}_s$' and `$\tilde{\mathbf{J}}_s$' to denote supercurrents in the twistor and momentum superspace, respectively, while simply `$\mathbf{J}_s$' for supercurrents in position superspace.

\subsection*{Conventions}
The spacetime vector indices are labeled with Greek alphabets (e.g. `${\mu}$'), with the flat Minkowski metric given by
\begin{align}
\eta_{{\mu}\nu}=\textbf{diag}(-1,1,1).
\end{align}
The spinor indices are denoted with lowercase Latin alphabets (e.g. `$a$'), where the raising and lowering of spinor indices is performed using the two dimensional Levi-Civita symbol
\begin{align}
A^a=\epsilon^{ab}A_b~,~A_b=\epsilon_{ab}A^a.
\end{align}
The spinor contractions are defined as follows
\begin{align}
A\cdot B:=\langle A B\rangle=A_a B^a.
\end{align}
The SO($\mathcal{N}$) R-symmetry indices are denoted by uppercase Latin alphabets (e.g. `$A$'), and are contracted using $\delta_{AB}$.

\section{Identities in position superspace}\label{app:2}
We present some important position superspace identities in this appendix. For $\mathcal{N}=1$ 
\begin{align}\label{N1}
(\tilde{X}_{12})_a^b&=(x_{12})_a^b+\frac{i}{2}(\theta_{1a}\theta_{2}^b+\theta_{1}^b\theta_{2a}).\\
(\tilde{X}_{12}^\pm)_a^b&=(\tilde{X}_{12})_a^b\pm\frac{i}{2}(\theta_{1a}-\theta_{2a})(\theta_{1}^b-\theta_{2}^b).\\
(\tilde{X}^{-}_{12})_a^c(\tilde{X}^{+}_{12})_c^b&=\tilde{x}_{12}^2\delta_a^b.\\
[(\tilde{X}_{ij}^\pm)_a^b]^T&=(\tilde{X}_{ij}^\mp)_b^a.\\
(\tilde{X}_{ij}^\pm)_a^b&=(\tilde{X}_{ij}^\mp)_a^b\pm i(\theta_{i,a}-\theta_{j,a})(\theta_{i}^b-\theta_{j}^b).\\
\theta_m\theta_n&=\frac{1}{2}\epsilon_{mn}\theta^2,\qquad\theta^m\theta^n=\frac{1}{2}\epsilon^{mn}\theta^2.
\end{align}
For $\mathcal{N}>1$, there are minor modifications
\begin{align}\label{N}
(\tilde{X}_{12})_a^b&=(x_{12})_a^b+\frac{i}{2}\delta_{AB}(\theta_{1a}^A\theta_{2}^{b,B}+\theta_{1}^{b,A}\theta_{2a}^B).\\
(\tilde{X}_{12}^\pm)_a^b&=(\tilde{X}_{12})_a^b\pm\frac{i}{2}\delta_{AB}(\theta_{1a}^A-\theta_{2a}^B)(\theta_{1}^{b,B}-\theta_{2}^{b,B}).\\
(\tilde{X}^{-}_{12})_a^c(\tilde{X}^{+}_{12})_c^b&=\bar{x}_{12}^2\delta_a^b.\\
[(\tilde{X}_{ij}^\pm)_a^b]^T&=(\tilde{X}_{ij}^\mp)_b^a.\\
(\tilde{X}_{ij}^\pm)_a^b&=(\tilde{X}_{ij}^\mp)_a^b\pm i\delta_{AB}(\theta_{i,a}^A-\theta_{j,a}^A)(\theta_{i}^{b,B}-\theta_{j}^{b,B}).
\end{align}


\subsection*{Projective vs non-projective integrals}
Recall the remaining integrals in \eqref{2ptCFTmid}. We stated that combining the projective measure $D\lambda_2$ with the Schwinger integral measure $dc_{12}$ gives the non-projective measure $d^2\lambda_2$. We now outline the derivation here
\begin{align}
&\int_{\mathbb{RP}^1}D\lambda_1D\lambda_2\int dc_{12}\;\textrm{Sgn}(c_{12})\textrm{exp}(-ic_{12}\lambda_{1a}\lambda_{2b}(x_{12})^{ab})=\int_{\mathbb{RP}^1}D\lambda_1\int d^2\lambda_2\;\textrm{exp}(-i\lambda_{1a}\lambda_{2b}(x_{12})^{ab})\notag\\
&=\int_{\mathbb{RP}^1}D\lambda_1\;\delta^{(2)}(\lambda_{1a}(x_{12})^{ab})
=\frac{1}{\textrm{Det}|(x_{12})^{ab}|}\int_{\mathbb{RP}^1}D\lambda_1\;\delta^{(2)}(\lambda_{1a})=\frac{1}{2x_{12}^2}.
\end{align}
Here, the final projective integral can be thought of as
\begin{align}
\int_{\mathbb{RP}^1}D\lambda_1\;\delta^{(2)}(\lambda_{1a})=\frac{\int d^{2}\lambda_1\;\delta^{(2)}(\lambda_{1a})}{\int \frac{dc}{|c|} }=\frac{\int d^{2}\lambda_1\;\delta^{(2)}(\lambda_{1a})}{\textrm{Vol(GL(1,$\mathbb{R}$))}}.
\end{align}
The projective and non-projective integrals are very closely related. For a function of correct homogeneity (for our case $r^{2s+2}$), the projective integral agrees with the non-projective one up to numerical factors.\footnote{For functions with wrong homogeneity, the projective integral is ill-defined, while the non-projective one evaluates to zero, as argued in \cite{Neiman:2017mel}.} Essentially, the extra one dimensional integral in the non-projective case leads to a factor of {Vol(GL(1;$\mathbb{R}$))}, which can be dealt by gauge-fixing the spinor $\lambda^a=(\lambda^1,\lambda^2)$ to $\lambda^a=(1,\lambda^2)$ \cite{Bala:2025qxr}.

\section{Spinor-helicity identities}\label{app:3}
We present some crucial spinor-helicity identities needed throughout this work. A detailed list of useful spinor identities can be found in \cite{Jain:2023idr}.

Any two-component spinor can be decomposed as a linear combination of $\lambda_{i}$ and $\bar{\lambda}_{i}$ using the following Schouten identities
\begin{align}\label{Schouten}
    \lambda_{i}^a&=-\frac{\langle i\bar{j}\rangle}{2p_{j}}\lambda_{j}^a+\frac{\langle ij\rangle}{2p_{j}}\bar{\lambda}_{j}^a,\notag\\
\bar{\lambda}_{i}^a&=-\frac{\langle\bar{i}\bar{j}\rangle}{2p_{j}}\lambda_{j}^a+\frac{\langle\bar{i}j\rangle}{2p_{j}}\bar{\lambda}_{j}^a.
\end{align}
The two dimensional delta function can be written in terms of one dimensional ones using
\begin{align}\label{Decomp}
\delta^{(2)}(k_1u_1^a+k_2u_2^a)=\frac{\delta(k_1)\delta(k_2)}{|u_1\cdot u_2|}.
\end{align}
\subsection*{Momentum-conserving delta function}
Recall that in \eqref{2ptWitten3}, we stated that the remaining delta functions neatly combined to produce the momentum-conserving delta function. Here we outline the derivation.
 \begin{align}
\delta^3(\vec{p_{1}}+\vec{p_{2}})&= \delta^3(p_{1}^{ab}+p_{2}^{ab}) = \delta^3\Big({\lambda_1^{(a}\bar{\lambda}_1^{b)}+\lambda_2^{(a}\bar{\lambda}_2^{b)}}\Big)\notag\\
&=\delta^3\left(\left(\frac{\langle \bar{1}2\rangle}{\langle 12\rangle}\right)\lambda_1^{a}\lambda_1^{b}+\left(\frac{\langle\bar{2}1\rangle}{\langle12\rangle}\right) \lambda_2^{a}\lambda_2^{b}+ \left(\frac{\langle\bar{1}1\rangle}{\langle12\rangle}-\frac{\langle \bar{2}2\rangle}{\langle 12\rangle}\right)\lambda_1^{a}\lambda_2^{b} \right)\notag\\ 
&=\frac{1}{|\text{Det}(\lambda_1^{a}\lambda_1^{b} \,\lambda_2^{c}\lambda_2^{d} \,\lambda_1^{e}\lambda_2^{f} )|} \delta\left(\frac{\langle\bar{2}1\rangle}{\langle12\rangle}\right)\delta\left(\frac{\langle \bar{1}2\rangle}{\langle 12\rangle}\right) \delta\left(\frac{\langle\bar{1}1\rangle}{\langle12\rangle}-\frac{\langle \bar{2}2\rangle}{\langle 12\rangle}\right)\notag\\
&=\frac{1}{|\langle 12\rangle|^3} \delta\left(\frac{\langle\bar{2}1\rangle}{\langle12\rangle}\right)\delta\left(\frac{\langle \bar{1}2\rangle}{\langle 12\rangle}\right) \delta\left(\frac{\langle\bar{1}1\rangle}{\langle12\rangle}-\frac{\langle \bar{2}2\rangle}{\langle 12\rangle}\right).
\end{align}
The momentum-conserving delta function for three point functions can be derived similarly.

\section{A detailed example of super-Penrose transform}\label{app:4}

In this appendix, we work out in detail an example of super-Penrose transform for three point functions. Consider the supercorrelator that involves a conserved spin-1 supercurrent along with two scalar supermultiplets of $\Delta=1$. Its twistor superspace expression is
\begin{align}\label{JJO}
&\langle 0| \mathbf{\hat{J}}(\lambda_1,\bar{\mu}_1,\psi_1)\mathbf{\hat{J}}_0(\lambda_2,\bar{\mu}_2,\psi_2)\mathbf{\hat{J}}_0(\lambda_3,\bar{\mu}_3,\psi_3)| 0\rangle \notag\\
&=\delta^{[1]}(\lambda_1\cdot\bar{\mu}_2-\lambda_2\cdot\bar{\mu}_1-\psi_1\psi_2)\delta^{[-1]}(\lambda_2\cdot\bar{\mu}_3-\lambda_3\cdot\bar{\mu}_2-\psi_2\psi_3)\delta^{[1]}(\lambda_3\cdot\bar{\mu}_1-\lambda_1\cdot\bar{\mu}_3-\psi_3\psi_1).
\end{align}
Imposing the incidence relations \eqref{IncidenceSUSY} and performing some simple manipulations results in
\begin{align}\label{JOOMid}
&\langle 0| \hat{\mathbf{J}}(\lambda_1,x_1,\theta_1)\hat{\mathbf{J}}_0(\lambda_2,x_2,\theta_2)\hat{\mathbf{J}}_0(\lambda_3,x_3,\theta_3)| 0\rangle=\delta^{[1]}(\lambda_{1a}\lambda_{2b}(\tilde{X}_{12}^{-})^{ab})\delta^{[-1]}(\lambda_{2c}\lambda_{3d}(\tilde{X}_{23}^{-})^{cd})\delta^{[1]}(\lambda_{3e}\lambda_{1f}(\tilde{X}_{31}^{-})^{ef}).
\end{align}
The super-Penrose transform for \eqref{JOOMid}, after contracting with polarization spinors, is
\begin{align}
&\langle 0| {\mathbf{J}}(x_1,\theta_1){\mathbf{J}}_0(x_2,\theta_2){\mathbf{J}}_0(x_3,\theta_3)| 0\rangle\notag\\
&=\int_{\mathbb{RP}^1}D\lambda_1D\lambda_2D\lambda_3\;(\xi_1\cdot\lambda_1)^2\;\langle 0| \mathbf{\hat{J}}(\lambda_1,\bar{\mu}_1,\psi_1)\mathbf{\hat{J}}_0(\lambda_2,\bar{\mu}_2,\psi_2)\mathbf{\hat{J}}_0(\lambda_3,\bar{\mu}_3,\psi_3)| 0\rangle\bigg\rvert_\mathfrak{X}.
\end{align}
The supercorrelator \eqref{JOOMid} can be expressed as a Schwinger integral, which takes the following form
\begin{align}\label{JOONPenrose}
&\langle 0| {\mathbf{J}}(x_1,\theta_1){\mathbf{J}}_0(x_2,\theta_2){\mathbf{J}}_0(x_3,\theta_3)| 0\rangle\notag\\
&=\int_{\mathbb{RP}^1}D\lambda_1D\lambda_2D\lambda_3\;(\xi_1\cdot\lambda_1)^2\int dc_{12}dc_{23}dc_{31}\frac{ic_{12}c_{31}}{c_{23}}\textrm{exp}(-ic_{12}\lambda_{1a}\lambda_{2b}(\tilde{X}_{12}^{-})^{ab}+\textrm{cyclic)}.
\end{align}
The integral in \eqref{JOONPenrose} can be solved by using partial integrations
\begin{align}
&\langle 0| {\mathbf{J}}(x_1,\theta_1){\mathbf{J}}_0(x_2,\theta_2){\mathbf{J}}_0(x_3,\theta_3)| 0\rangle\notag\\
&=\int_{\mathbb{RP}^1}D\lambda_1D\lambda_2D\lambda_3\;(\xi_1\cdot\lambda_1)\int dc_{12}dc_{23}dc_{31}\frac{c_{31}}{c_{23}}\textrm{exp}(-ic_{23}\lambda_{2c}\lambda_{3d}(\tilde{X}_{23}^{-})^{cd})\cross\notag\\
&\Big(\xi_{1m}\frac{(\tilde{X}^{-}_{12})^{mn}}{\tilde{x}_{12}^2}\frac{\partial}{\partial \lambda_2^n}\textrm{exp}(-ic_{12}\lambda_{1a}\lambda_{2b}(\tilde{X}_{12}^{-})^{ab})\Big)\textrm{exp}(-ic_{31}\lambda_{3e}\lambda_{1f}(\tilde{X}_{31}^{-})^{ef})\notag\\
&=\textrm{TD}-\int_{\mathbb{RP}^1}D\lambda_1D\lambda_2D\lambda_3\;(\xi_1\cdot\lambda_1)\int dc_{12}dc_{23}dc_{31}\frac{c_{31}}{c_{23}}\textrm{exp}(-ic_{12}\lambda_{1a}\lambda_{2b}(\tilde{X}_{12}^{-})^{ab})\cross\notag\\
&\Big(\xi_{1m}\frac{(\tilde{X}^{-}_{12})^{mn}}{\tilde{x}_{12}^2}\frac{\partial}{\partial \lambda_2^n}\textrm{exp}(-ic_{23}\lambda_{2c}\lambda_{3d}(\tilde{X}_{23}^{-})^{cd})\Big)\textrm{exp}(-ic_{31}\lambda_{3e}\lambda_{1f}(\tilde{X}_{31}^{-})^{ef})\notag\\
&=-\int_{\mathbb{RP}^1}D\lambda_1D\lambda_2D\lambda_3\;(\xi_1\cdot\lambda_1)\int dc_{12}dc_{23}dc_{31}{ic_{31}}\Big(\frac{\xi_{1m}(\tilde{X}^{+}_{12})^{m}_{n}(\tilde{X}^{+}_{23})^{n}_{o}\lambda_{3}^o}{\tilde{x}_{12}^2}\Big)\cross\notag\\
&\textrm{exp}(-ic_{12}\lambda_{1a}\lambda_{2b}(\tilde{X}_{12}^{-})^{ab})\textrm{exp}(-ic_{23}\lambda_{2c}\lambda_{3d}(\tilde{X}_{23}^{-})^{cd})\textrm{exp}(-ic_{31}\lambda_{3e}\lambda_{1f}(\tilde{X}_{31}^{-})^{ef})\notag\\
&=\int_{\mathbb{RP}^1}D\lambda_1D\lambda_2D\lambda_3\;(\xi_1\cdot\lambda_1)\Big(\frac{\xi_{1m}(\tilde{X}^{+}_{12})^{m}_{n}(\tilde{X}^{+}_{23})^{n}_{o}\lambda_{3}^o}{\tilde{x}_{12}^2}\Big)\delta(\lambda_{1a}\lambda_{2b}(\tilde{X}_{12}^{-})^{ab})\delta(\lambda_{2c}\lambda_{3d}(\tilde{X}_{23}^{-})^{cd})\delta^{[1]}(\lambda_{3e}\lambda_{1f}(\tilde{X}_{31}^{-})^{ef}),
\end{align}
where we have used the fact that $[(\tilde{X}_{ij}^\pm)_a^b]^T=(\tilde{X}_{ij}^\mp)_b^a$. At this stage, we projectively integrate over $\lambda_1\And\lambda_2$ using \eqref{Delta}, which leads to the following expression
\begin{align}\label{JOOLast}
&\langle 0| {\mathbf{J}}(x_1,\theta_1){\mathbf{J}}_0(x_2,\theta_2){\mathbf{J}}_0(x_3,\theta_3)| 0\rangle\notag\\
&=\int_{\mathbb{RP}^1}D\lambda_3\big(\xi_{1p}(\tilde{X}^{+}_{12})^{p}_{q}(\tilde{X}^{+}_{23})^{q}_{r}\lambda_3^r\big)\Big(\frac{\xi_{1m}(\tilde{X}^{+}_{12})^{m}_{n}(\tilde{X}^{+}_{23})^{n}_{o}\lambda_{3}^o}{\tilde{x}_{12}^2}\Big)\delta^{[1]}_{Sym}(\lambda_{3a}(\tilde{X}_{31}^{+})^{a}_b(\tilde{X}_{12}^{+})^{b}_c(\tilde{X}_{23}^{+})^{c}_d\lambda_{3}^d),
\end{align}
where we manifestly symmetrize the argument in the remaining delta function using \eqref{Sym}.\\
We now insert the appropriate inverse operator \eqref{InverseOp} to solve this integral.
\begin{align}
&\langle 0| {\mathbf{J}}(x_1,\theta_1){\mathbf{J}}_0(x_2,\theta_2){\mathbf{J}}_0(x_3,\theta_3)| 0\rangle=\int_{\mathbb{RP}^1}D\lambda_3\big(\xi_{1p}(\tilde{X}^{+}_{12})^{p}_{q}(\tilde{X}^{+}_{23})^{q}_{r}\lambda_3^r\big)\Big(\xi_{1m}\frac{(\tilde{X}^{+}_{12})^{m}_{n}(\tilde{X}^{+}_{23})^{n}_{o}}{\tilde{x}_{12}^2}\Big)\cross\notag\\
&
\Bigg(\frac{(\tilde{X}_{31}^{+})^{o}_i(\tilde{X}_{12}^{+})^{i}_j(\tilde{X}_{23}^{+})^{j}_k+(\tilde{X}_{23}^{-})^{o}_i(\tilde{X}_{12}^{-})^{i}_j(\tilde{X}_{31}^{-})^{j}_k}{2\;\tilde{x}_{12}^2\tilde{x}_{23}^2\tilde{x}_{31}^2}\Bigg)\frac{\partial}{\partial \lambda_{3k}}\delta_{Sym}\big(\lambda_{3a}(\tilde{X}_{31}^{+})^{a}_b(\tilde{X}_{12}^{+})^{b}_c(\tilde{X}_{23}^{+})^{c}_d\lambda_{3}^d)
\end{align}
Performing the partial integration in \eqref{JOOLast} leads to the following expression
\begin{align}
&\langle 0| {\mathbf{J}}(x_1,\theta_1){\mathbf{J}}_0(x_2,\theta_2){\mathbf{J}}_0(x_3,\theta_3)| 0\rangle
=\int_{\mathbb{RP}^1}D\lambda_3\;\delta_{Sym}\big(\lambda_{3a}(\tilde{X}_{31}^{+})^{a}_b(\tilde{X}_{12}^{+})^{b}_c(\tilde{X}_{23}^{+})^{c}_d\lambda_{3}^d)\big(\xi_{1p}(\tilde{X}^{+}_{12})^{p}_{q}(\tilde{X}^{+}_{23})^{q}_{r}\lambda_3^r\big)\cross\notag\\
&\frac{\xi_{1m}(\tilde{X}^{+}_{12})^{m}_{n}(\tilde{X}^{+}_{23})^{n}_{o}}{\tilde{x}_{12}^2}\Bigg(\frac{(\tilde{X}_{31}^{+})^{o}_i(\tilde{X}_{12}^{+})^{i}_j(\tilde{X}_{23}^{+})^{j}_k+(\tilde{X}_{23}^{-})^{o}_i(\tilde{X}_{12}^{-})^{i}_j(\tilde{X}_{31}^{-})^{j}_k}{2\;\tilde{x}_{12}^2\tilde{x}_{23}^2\tilde{x}_{31}^2}\Bigg)\frac{\partial}{\partial \lambda_{3k}}\big(\xi_{1p}(\tilde{X}^{-}_{12})^{p}_{q}(\tilde{X}^{-}_{23})^{q}_{r}\lambda_3^r\big).
\end{align}
Performing partial integration and simplifying leads to the following expression
\begin{align}\label{JOOSF}
\langle 0| {\mathbf{J}}(x_1,\theta_1){\mathbf{J}}_0(x_2,\theta_2){\mathbf{J}}_0(x_3,\theta_3)| 0\rangle&=\frac{\xi_{1p}(\tilde{X}_{12}^{-})^{p}_q(\tilde{X}_{23}^{-})^{q}_r(\tilde{X}_{31}^{-})^{r}_s\xi_1^{s}}{\tilde{x}_{12}^2\tilde{x}_{31}^2}\int_{\mathbb{RP}^1}D\lambda_3\;\delta_{Sym}\big(\lambda_{3a}(\tilde{X}_{31}^{+})^{a}_b(\tilde{X}_{12}^{+})^{b}_c(\tilde{X}_{23}^{+})^{c}_d\lambda_{3}^d\big).
\end{align}
After performing the final integral using \eqref{Det}, the expression \eqref{JOOSF} is compactly written in terms of the superconformal building block
\begin{align}\label{JOOFinal}
\langle 0| {\mathbf{J}}(x_1,\theta_1){\mathbf{J}}_0(x_2,\theta_2){\mathbf{J}}_0(x_3,\theta_3)| 0\rangle&=\frac{{Q}_1}{\tilde{x}_{12}\tilde{x}_{23}\tilde{x}_{31}},
\end{align}
where
\begin{align}
Q_1=\frac{\xi_{1p}(\tilde{X}_{12}^{-})^{p}_q(\tilde{X}_{23}^{-})^{q}_r(\tilde{X}_{31}^{-})^{r}_s\xi_1^{s}}{\tilde{x}_{12}^2\tilde{x}_{31}^2}.
\end{align}
This is the correct three point function as mentioned in \cite{Nizami:2013tpa}.

\section{A detailed example of super-Witten transform}\label{app:5}
In this appendix, we work out the super-Witten transform in detail. Recall \eqref{3ptSW} in the $\mathcal{N}=1$ case. Plugging in the integral representation of the supercorrelator leads to
\begin{align}
&\langle0|\tilde{\textbf{J}}_{s_1}(\lambda_1,\bar{\lambda}_1,\eta_1,\bar{\eta}_1)\tilde{\textbf{J}}_{s_2}(\lambda_2,\bar{\lambda}_2,\eta_2,\bar{\eta}_2)\tilde{\textbf{J}}_{s_3}(\lambda_3,\bar{\lambda}_3,\eta_3,\bar{\eta}_3)|0\rangle\notag\\
=&\prod_{i=1}^3\int d^2\bar{\mu}_i\;\textrm{exp}(-i\bar{\lambda}_{ia}\bar{\mu}_i^a)\int d\chi_i\;\textrm{exp}\Big(\frac{\chi_i\bar{\eta}_i}{4}\Big)(\chi_i-\eta_i)\int dc_{12}dc_{23}dc_{31}\;c_{12}^{s_1+s_2-s_3}c_{23}^{s_2+s_3-s_1}c_{31}^{s_3+s_1-s_2}\notag\\
&(1-ic_{12}\psi_1\psi_2)(1-ic_{23}\psi_2\psi_3)(1-ic_{31}\psi_3\psi_1)\;\textrm{exp}(i(c_{12}(\lambda_1\cdot\bar{\mu}_2-\lambda_2\cdot\bar{\mu}_1)+\textrm{cyclic}))\bigg\rvert_{\mathfrak{P}}.
\end{align}
Combining all the exponentials involving the bosonic spinors and integrating over $\bar{\mu}_i$ results in the following expression
\begin{align}
&\langle0|\tilde{\textbf{J}}_{s_1}(\lambda_1,\bar{\lambda}_1,\eta_1,\bar{\eta}_1)\tilde{\textbf{J}}_{s_2}(\lambda_2,\bar{\lambda}_2,\eta_2,\bar{\eta}_2)\tilde{\textbf{J}}_{s_3}(\lambda_3,\bar{\lambda}_3,\eta_3,\bar{\eta}_3)|0\rangle\notag\\
=&\int dc_{12}dc_{23}dc_{31}\;\delta^{(2)}(\bar{\lambda}_1^a+c_{12}\lambda_2^a-c_{31}\lambda_3^a)\delta^{(2)}(\bar{\lambda}_2^a+c_{23}\lambda_3^a-c_{12}\lambda_1^a)\delta^{(2)}(\bar{\lambda}_3^a+c_{31}\lambda_1^a-c_{23}\lambda_2^a)\notag\\
&\prod_{i=1}^3\int d\chi_i\;\textrm{exp}\Big(\frac{\chi_i\bar{\eta}_i}{4}\Big)(\chi_i-\eta_i)
(1-ic_{12}\psi_1\psi_2)(1-ic_{23}\psi_2\psi_3)(1-ic_{31}\psi_3\psi_1)c_{12}^{s_1+s_2-s_3}c_{23}^{s_2+s_3-s_1}c_{31}^{s_3+s_1-s_2} \bigg\rvert_{\mathfrak{P}}.
\end{align}
We now use the Schouten identities \eqref{Schouten} to convert $\bar{\lambda}_i$'s to $\lambda_i$'s, and then convert the delta functions using \eqref{Decomp}. This yields the following form
\begin{align}
&\langle0|\tilde{\textbf{J}}_{s_1}(\lambda_1,\bar{\lambda}_1,\eta_1,\bar{\eta}_1)\tilde{\textbf{J}}_{s_2}(\lambda_2,\bar{\lambda}_2,\eta_2,\bar{\eta}_2)\tilde{\textbf{J}}_{s_3}(\lambda_3,\bar{\lambda}_3,\eta_3,\bar{\eta}_3)|0\rangle\notag\\
=&\int dc_{12}dc_{23}dc_{31}\;\frac{1}{|\langle12\rangle\langle23\rangle\langle31\rangle|}\delta\left(c_{12}+\frac{\langle\bar{1}3\rangle}{\langle23\rangle}\right)\delta\left(c_{23}+\frac{\langle\bar{2}1\rangle}{\langle31\rangle}\right)\delta\left(c_{31}+\frac{\langle\bar{3}2\rangle}{\langle12\rangle}\right)\notag\\
&\delta\left(c_{12}-\frac{\langle\bar{2}3\rangle}{\langle13\rangle}\right)\delta\left(c_{23}-\frac{\langle\bar{3}1\rangle}{\langle21\rangle}\right)\delta\left(c_{31}-\frac{\langle\bar{1}2\rangle}{\langle32\rangle}\right)c_{12}^{s_1+s_2-s_3}c_{23}^{s_2+s_3-s_1}c_{31}^{s_3+s_1-s_2}\notag\\
&\prod_{i=1}^3\int d\chi_i\;\textrm{exp}(\frac{\chi_i\bar{\eta}_i}{4})(\chi_i-\eta_i))
(1-ic_{12}\psi_1\psi_2)(1-ic_{23}\psi_2\psi_3)(1-ic_{31}\psi_3\psi_1) \bigg\rvert_{\mathfrak{P}}.
\end{align}
At this stage, we perform the integral over the Schwinger parameters. Simplifying the angle brackets using spinor identities leads to the following expression
\begin{align}
&\langle0|\tilde{\textbf{J}}_{s_1}(\lambda_1,\bar{\lambda}_1,\eta_1,\bar{\eta}_1)\tilde{\textbf{J}}_{s_2}(\lambda_2,\bar{\lambda}_2,\eta_2,\bar{\eta}_2)\tilde{\textbf{J}}_{s_3}(\lambda_3,\bar{\lambda}_3,\eta_3,\bar{\eta}_3)|0\rangle\notag\\
=&\prod_{m=1}^3\int d\chi_m\;\textrm{exp}\Big(\frac{\chi_m\bar{\eta}_m}{4}\Big)(\chi_m-\eta_m)\prod_{\substack{i,j,k=1 \\ i<j\\k\neq i,j}}^3\left(1-\frac{(\chi_i+\eta_i)(\chi_j+\eta_j)}{8}\frac{\langle\bar{i}\bar{j}\rangle}{E}\right)\left(\frac{\langle\bar{i}\bar{j}\rangle}{E}\right)^{s_i+s_j-s_k}\cross\notag\\
&\Bigg(\frac{1}{|\langle12\rangle\langle23\rangle\langle31\rangle|}\delta\left(\frac{\langle\bar{2}3\rangle}{\langle13\rangle}+\frac{\langle\bar{1}3\rangle}{\langle23\rangle}\right)\delta\left(\frac{\langle\bar{3}1\rangle}{\langle21\rangle}+\frac{\langle\bar{2}1\rangle}{\langle31\rangle}\right)\delta\left(\frac{\langle\bar{1}2\rangle}{\langle32\rangle}+\frac{\langle\bar{3}2\rangle}{\langle12\rangle}\right)\Bigg),
\end{align}
where $E=p_1+p_2+p_3$. The terms in the final line combine to give the momentum-conserving delta functions. Finally, performing the Grassmann integrals results in
\begin{align}
&\langle0|\tilde{\textbf{J}}_{s_1}(\lambda_1,\bar{\lambda}_1,\eta_1,\bar{\eta}_1)\tilde{\textbf{J}}_{s_2}(\lambda_2,\bar{\lambda}_2,\eta_2,\bar{\eta}_2)\tilde{\textbf{J}}_{s_3}(\lambda_3,\bar{\lambda}_3,\eta_3,\bar{\eta}_3)|0\rangle\notag\\
&=\frac{\langle\bar{1}\bar{2}\rangle^{s_1+s_2-s_3}\langle\bar{2}\bar{3}\rangle^{s_2+s_3-s_1}\langle\bar{3}\bar{1}\rangle^{s_3+s_1-s_2}}{E^{s_1+s_2+s_3}}\textrm{exp}\left(\frac{\eta_1\bar{\eta}_1+\eta_2\bar{\eta}_2+\eta_3\bar{\eta}_3}{4}\right)\left(1-\eta_1\eta_2\frac{\langle\bar{1}\bar{2}\rangle}{2E}-\eta_2\eta_3\frac{\langle\bar{2}\bar{3}\rangle}{2E}-\eta_3\eta_1\frac{\langle\bar{3}\bar{1}\rangle}{2E}\right),
\end{align}
which is the three point supercorrelator as mentioned in \eqref{JJJ}.
\newpage
\bibliographystyle{JHEP}
\bibliography{biblio}
\end{document}